\documentclass[11pt,a4paper]{article}

\pdfoutput=1
\usepackage{jcappub}

\usepackage[english]{babel}  
\usepackage[T1]{fontenc}
\usepackage[load-configurations = abbreviations]{siunitx}
\usepackage{url}
\usepackage{xfrac}
\usepackage[dvipsnames]{xcolor}
\usepackage{float}
\usepackage{multirow}
\usepackage{subcaption}
\usepackage{bm}
\usepackage{ragged2e}

\hypersetup{urlcolor=RedViolet,
	    citecolor=ForestGreen ,
	    linkcolor=RoyalBlue, 
	    colorlinks=true
}

\let\vec\mathbf 

\title{\huge\justifying Directionality and head-tail recognition in the keV-range with the MIMAC detector by deconvolution of the ionic signal}

\author[a]{Cyprien Beaufort,}
\author[a]{Olivier Guillaudin,}
\author[a]{Jean-François Muraz,}
\author[a]{Nadine Sauzet,}
\author[a]{Daniel Santos,}
\author[b]{Richard Babut}

\affiliation[a]{Laboratoire de Physique Subatomique et de Cosmologie, Univ. Grenoble-Alpes (UGA), CNRS, Grenoble INP$^*$, LPSC-IN2P3, 38000 Grenoble, France}
\affiliation[*]{Institute of Engineering Univ. Grenoble Alpes}
\affiliation[b]{Institut de Radioprotection et de Sûreté Nucléaire (IRSN), BP 3, 13115, Saint-Paul-lez-Durance, Cedex, France}

\emailAdd{beaufort@lpsc.in2p3.fr}
\emailAdd{guillaudin@lpsc.in2p3.fr}
\emailAdd{muraz@lpsc.in2p3.fr}
\emailAdd{sauzet@lpsc.in2p3.fr}
\emailAdd{santos@lpsc.in2p3.fr}
\emailAdd{richard.babut@irsn.fr}

\abstract{
Directional detection is the only strategy for the unambiguous identification of galactic Dark Matter (DM) even in the presence of an irreducible background such as beyond the neutrino floor. This approach requires measuring the direction of a DM-induced nuclear recoil in the keV-range. To probe such low energies, directional detectors must operate at high gain where 3D track reconstruction can be distorted by the influence of the numerous ions produced in the avalanches. The article describes the interplay between electrons and ions during signal formation in a Micromegas. It introduces \texttt{SimuMimac}, a simulation tool dedicated to high gain detection that agrees with MIMAC measurements. This work proposes an analytical formula to deconvolve the ionic signal induced on the grid from any measurements, with no need for prior nor \textit{ad hoc} parameter. This deconvolution is experimentally tested and validated, revealing the fine structure of the primary electrons cloud and consequently leading to head-tail recognition in the keV-range. Finally, the article presents how this deconvolution can be used for directionality by reconstructing the spectra of mono-energetic $27\,\mathrm{keV}$ and $8\,\mathrm{keV}$ neutrons with an angular resolution better than $15^\circ$. This novel approach for directionality appears as complementary to the standard one from 3D tracks reconstruction and offers redundancy for improving directional performances at high gain in the keV region.
}

\arxivnumber{}

\begin{document}

\maketitle

\section{Introduction}

The direct detection of a Dark Matter (DM) particle represents one of the major challenges in particle physics \cite{Billard2021}. The Weakly Interacting Massive Particle (WIMP) acts as a leading non-baryonic DM candidate. While the main direct detection projects keep improving their sensitivity \cite{Schumann2019}, they will soon reach the neutrino floor beyond which a WIMP signal cannot be distinguished from the background \cite{Billard2013}. They also encounter an issue to claim for a WIMP detection unambiguously disentangled from the environmental background \cite{Mayet2016}. Directional detection aims to cope with those two obstacles by making use of the unique signature of galactic DM. A DM-nucleus scattering will induce a nuclear recoil with an anisotropic angular distribution correlated with the Earth's motion \cite{Catena2015, Kavanagh2015}. The directional detection strategy relies on the simultaneous measurements of the energy and the direction of a DM-induced nuclear recoil for the identification of a DM particle without ambiguity \cite{Mayet2016, Vahsen2021}. This directional information is generally measured in gaseous detectors, from the ionization signal \cite{Riffard2015, Ikeda2021} or via the optical readout \cite{Baracchini2020, Amaro2022}, or by exploiting nuclear emulsions in solid detectors \cite{Gorbunov2020}. The complete reconstruction of the direction of a nuclear recoil also requires to discriminate between the head and the tail of the track, usually established through measurements of a charge asymmetry \cite{Burgos2008}.

Recoil energies must be searched in the keV-range: a WIMP typically transfers at maximum an energy lower than $10\,\mathrm{keV/nucleon}$ \cite{Sciolla2009}. Accessing directionality in this low energy region turns out to be experimentally challenging \cite{Battat2016}. In order to fully describe the nuclear recoil track at low energy, directional detectors must be sensitive to any primary charge which requires to operate at high gain (above $10^4$). Recently we have published measurements of fluorine ions down to $6.3\,\mathrm{keV}$ \cite{Tao2019, Tao2020} highlighting some issues with high gain detection: we observed an elongation of the signal duration and a distortion of the 3D track reconstruction, both being correlated to the detector gain. The current work proposes an explanation for these measurements and validates it through simulations, analytical calculations, and experimental data. 

In this paper, we describe the low-energy performances of MIMAC \cite{Santos2013} using a Micromegas \cite{Giomataris1996} for directional detection. At high gain, the detector gets more sensitivity to the signal induced by the numerous ions accumulated in the amplification area. The interplay between the abrupt electronic signal and the blunt ionic one causes the measured distortions of the 3D tracks. In Section \ref{sec:sigForm} we introduce the MIMAC detector and we illustrate how the ionic signal influences the detection. We have developed a simulation tool to numerically model the signal formation in MIMAC at high gain. The simulations, described in Section \ref{sec:SimuMimac}, are in agreement with the previously published results and have enabled to rule out the hypothesis of an event-based space-charge built at high gain. 

The core of the article lies in the deconvolution of the signal induced by the ions on the grid of the Micromegas. The separation between the ionic and the electronic signals leads to deep improvements in detection. First, it fixes the issue of gain-dependent track elongations. Second, it reveals the fine structure of the electronic current and makes use of its fast kinematics to access a precise time distribution of the primary electrons cloud. Section \ref{sec:Prim} is dedicated to experimental tests of an analytical formula for deconvolving the ionic contribution. Finally, we propose to explore in Section \ref{sec:directionality} some of the new possibilities offered by the deconvolution in comparing the electronic and ionic signals in order to determine the scattering angle of a neutron-proton interaction and to evaluate the directional performances of MIMAC when operating at high gain. This section is considered as an exploratory window toward directional DM detection at high gain.

\section{Signal formation in the MIMAC detector}\label{sec:sigForm}

Before describing the MIMAC detector, we will briefly review the main physical processes involved in particle detection with a Micromegas detector, although the backbone of these principles is common to most gaseous detectors. A Micromegas placed into a Time Projection Chamber (TPC) \cite{Sauli1977} consists of a two-stage structure: (1) a drift area corresponding to the active volume in between a cathode and a grid; (2) an amplification area in between a grid and an anode. The length of the amplification area, called \textit{gap}, plays a crucial role in this work. Figure~\ref{fig:mimacChamber} shows an illustration of the detection principle of a TPC coupled with a Micromegas. A WIMP or a neutron entering the active volume could make an elastic collision on a nucleus of the gas, consequently inducing a nuclear recoil. The recoil releases its energy through three competing processes: ionization, scintillation, and heat production. The amount of kinetic energy released as ionization is called the Ionization Quenching Factor (IQF) and will be discussed in Section~\ref{sec:directionality}.

The electrons produced by ionization along the track of the nuclear recoil drift towards the grid due to the influence of a uniform electric field. The electrons suffer diffusion on the way, which enlarges the size of the primary electrons cloud. Once a primary electron reaches the grid, it enters the amplification area and it sufficiently accelerates to ionize the gas, leading to a Townsend avalanche. The secondary electrons are collected on the anode while the ions are collected on the grid. The motion of the secondary charges induces a current both on the anode and on the grid, described by the Ramo-Shockley theorem \cite{Ramo1939, Shockley1938}:
\begin{equation}
	i_k(t) = \sum_n q_n\, \vec{E}^w_k(\vec{r}_n)\cdot\vec{v}(\vec{r}_n)
	\label{eq:Ramo}
\end{equation}
where $k$ labels the electrode (grid or anode), the sum is over all the moving charges, and $\vec{v}(\vec{r}_n)$ is the velocity of the $n$-th moving charge. The weighting field $\vec{E}^w_k(\vec{r}_n)$ is defined as the electric field at coordinates $\vec{r}_n$ of the charge $n$, if the charge $n$ is removed and if all electrodes are grounded except the electrode $k$ whose potential is set to $1\,\mathrm{V}$. The interplay between the electronic signal and the ionic signal represents the central element under study in this paper.

	\subsection{The MIMAC detector}
	
	The MIMAC detector makes use of TPCs coupled to pixelated Micromegas in order to measure the ionization energy deposed by a particle and to reconstruct at the same time its track in 3D \cite{Santos2013}. MIMAC stands for MIcro-TPC MAtrix of Chambers since it can assemble several $\mu$-TPC chambers to achieve a large volume detection of $1\,\mathrm{m^3}$ or even more \cite{Couturier2016, Billard2012}. In this paper, we analyse data collected in a single MIMAC chamber.

\begin{figure*}
	\captionsetup[subfigure]{justification=centering}
	\centering
	\begin{minipage}{0.6\linewidth}	
		\includegraphics[width=\linewidth]{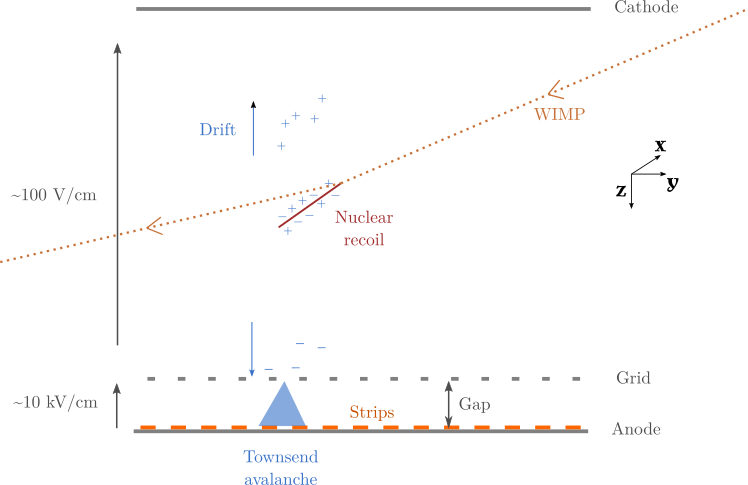}
		\subcaption{A MIMAC chamber: a TPC coupled\\with a pixelated Micromegas}
		\label{fig:mimacChamber}
	\end{minipage}
	\hfill
	\begin{minipage}{0.3\linewidth}	
		\vspace*{1cm} 
		\includegraphics[width=\linewidth]{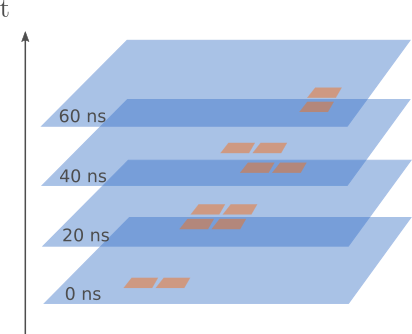}
		\vspace*{0.8cm} 
		\subcaption{\centering 3D track reconstruction from the anode}
		\label{fig:mimac3D}
	\end{minipage}
	\caption{Schematic of the MIMAC detection principle. A WIMP induces a nuclear recoil that ionizes the gas. The primary charges drift along the electric field lines direction, \textit{i.e.} along the Z-axis, towards the grid. They then enter the amplification area where they produce avalanches. The signals induced on the grid and on the pixelated anode are measured every $20~\rm{ns}$.}
	\label{fig:MimacSchem}
\end{figure*}	

The chamber is filled with a low-pressure gas mixture. In this paper we present measurements in 50\% i-C$_4$H$_{10}$ + 50\% CHF$_3$ at $30\,\mathrm{mbar}$ or in the \textit{Mimac gas} composed of 70\% CF$_4$ + 28\% CHF$_3$ + $2\%$ i-C$_4$H$_{10}$ at $50\,\mathrm{mbar}$. The voltages, the gas mixture, and the pressure are adjusted to meet the physical requirements for low-mass WIMP searches: high-gain, intermediate electron drift velocity, and long-enough protons tracks to reconstruct the 3D. For instance, in the experimental conditions of Section \ref{sec:directionality}, we have an electron drift velocity of $11.6\,\mathrm{\mu m / ns}$ (according to \texttt{Magboltz} \cite{Magboltz}) and track lengths of $3.2\,\mathrm{mm}$ for $10\,\mathrm{keV}$ protons (according to \texttt{SRIM} \cite{SRIM}). The proton represents an interesting target for low-mass WIMP searches since it is the lightest odd nucleus for spin dependent interactions. In MIMAC, one can easily change the gas mixture or the experimental conditions to adapt the detector to some specific particle searches. 

The MIMAC chamber has $25\,\mathrm{cm}$ of drift maintained at a uniform electric field of $86\,\mathrm{V/cm}$ thanks to a field cage. The electric field lines are oriented along the Z-axis of the detector. We are using a bulk Micromegas \cite{Giomataris2004} with a large amplification gap of $512\,\mathrm{\mu m}$ (dedicated to low pressure measurements) polarized at a high-voltage to produce the amplification field of order  $\mathcal{O}(10\,\mathrm{kV/cm})$. The Micromegas anode is pixelated and it contains $256$ strips in both X and Y directions with a pitch of $424.3\,\mathrm{\mu m}$, covering a total area of $10.8\times10.8\,\mathrm{cm^2}$ \cite{Iguaz2011}. A strip is fired when the current induced by the moving charges exceeds a threshold that is automatically calibrated before detector polarization in order to remain above the strip intrinsic noise \cite{Bourrion2011}. 
\begin{figure*}
	\centering
	\includegraphics[width=0.95\linewidth]{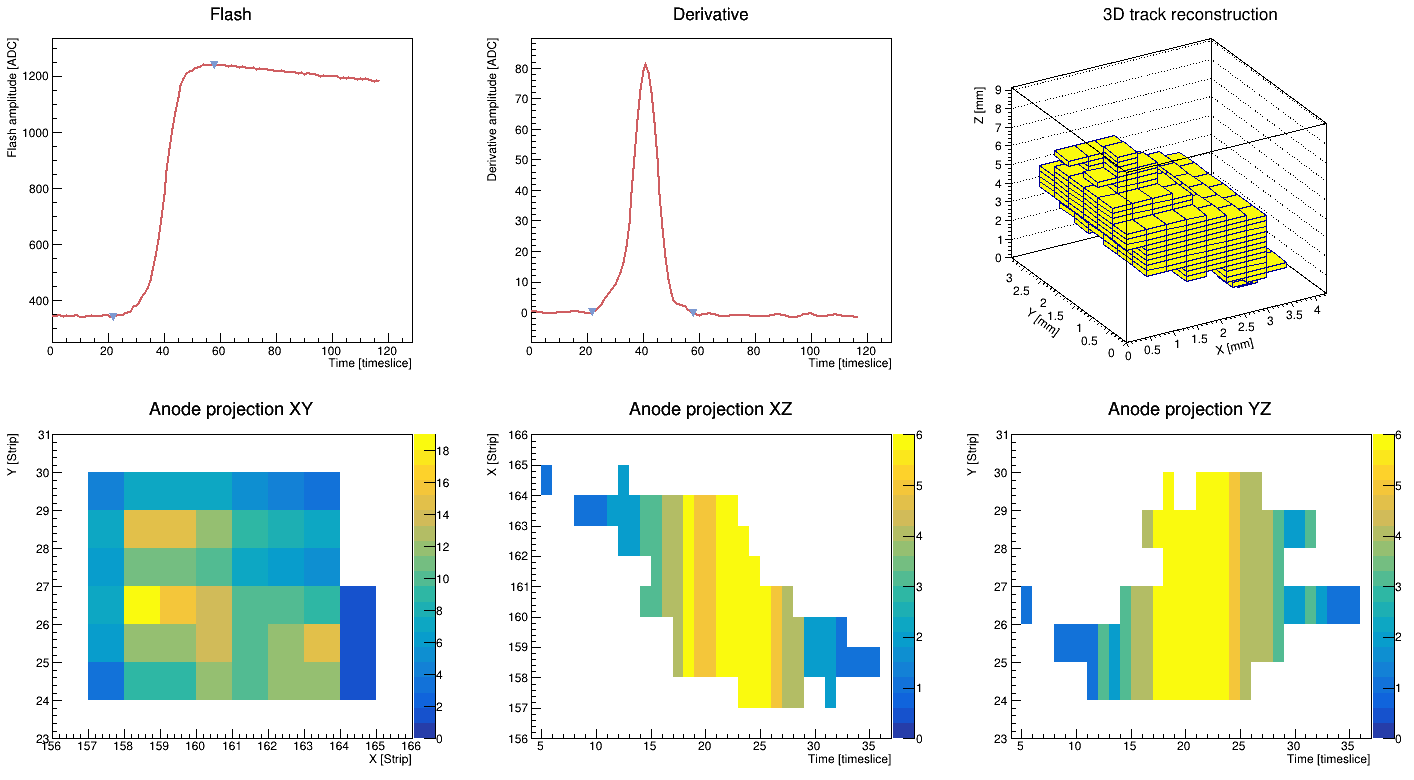}
	\caption{Example of a MIMAC measurement of a proton recoil of $8.6\,\mathrm{keV_{nr}}$ obtained by elastic scattering from a $27\,\mathrm{keV}$ neutron during the AMANDE campaign detailed in Section \ref{sec:directionality}. The Flash signal is presented in the upper left plot. The 3D track reconstruction on the upper right plot is obtained from the combination of the measurements on the pixelated anode presented in the lower panel of the figure.}
	\label{fig:recoilExample}
\end{figure*}

We measure simultaneously the signal induced on the grid and the one induced on the pixelated anode by two independent but synchronized readouts, both sampled at $50\,\mathrm{MHz}$ ($20\,\mathrm{ns}$) by a self-triggered electronic system. On the grid, we use a charge sensitive pre-amplifier that integrates the charge over time and which is digitalized via a Flash-ADC with one value per \textit{timeslice}, corresponding to a time sample of $20\,\mathrm{ns}$. This signal is later referred to as \textit{Flash}. We define the ionization energy (in ADC units) as the amplitude of the Flash signal which is proportional to the number of primary electrons. In parallel, a set of 8 specially designed 64-channel MIMAC ASIC \cite{Richer2011} reads the pixelated anode and measures in 2D the position of the activated strips for each timeslice of $20\,\mathrm{ns}$. 

As presented in Figure~\ref{fig:mimac3D}, the reconstruction of the track in 3D is performed by superimposing the 2D anode readouts for each timeslice. The Z-coordinate is provided by the combination of the constant electron drift velocity (estimated with \texttt{Magboltz}) and the time sampling. However, this reconstruction of the Z-coordinate only offers relative information and does not determine the absolute position of the interaction point along the Z-axis. While the MIMAC team has shown how to localize in 3D the interaction point thanks to the signal induced on the cathode \cite{Couturier2017}, this approach is not accessible yet in the low-energy range (below $30\,\mathrm{keV}$) considered in this paper. An example of a typical MIMAC measurement is presented in Figure~\ref{fig:recoilExample}.
	
	\subsection{The influence of the ionic signal at high gain}

	We have already demonstrated that MIMAC can access directionality and reconstruct a neutron spectrum at intermediate energies ($127\,\mathrm{keV}$ \cite{Maire2013} and $27\,\mathrm{keV}$ \cite{Maire2016}) when operating at low gain with a gap of $256\,\mathrm{\mu m}$. However, we are interested in accessing directionality in the keV-range for which operating at high gain ($>10^4$) is mandatory. For this reason, we use a large gap of $512\,\mathrm{\mu m}$. We have recently shown that in such conditions we suffer from an elongation of the reconstructed tracks, the elongation being correlated to the gain of the detector \cite{Tao2019}. By applying empirical correction we achieved a $15^\circ$ angular resolution for $>10\,\mathrm{keV}$ fluorine ions sent along the drift direction \cite{Tao2020}. The present work aims to demonstrate the origin of the measured elongation, how to deal with it, and how to use it to improve our directional performances.
	
	\begin{figure*}
	\centering
	\includegraphics[width=0.95\linewidth]{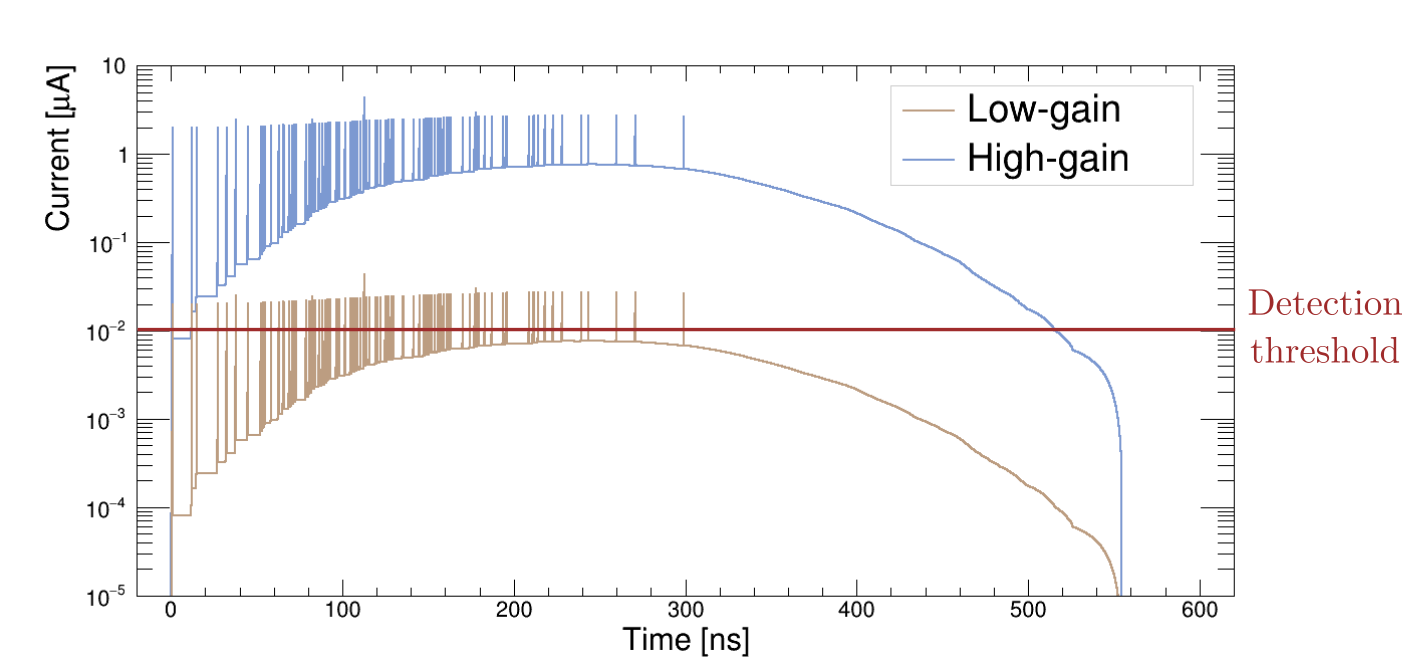}
	\caption{Influence of the gain on detection. The primary electrons cloud follows a typical distribution for a single event of $10\,\mathrm{keV}$ proton. The current is determined analytically from the model described in Section \ref{sec:Prim}. The peaks correspond to the electronic signal, with a typical duration of $1~\rm{ns}$, and the baseline is the contribution of the ionic signal.}
	\label{fig:sigFormation}
\end{figure*}
	
In the avalanche, most of the electron-ion pairs are produced close to the anode. All the electrons are collected in less than $1\,\mathrm{ns}$ whereas the ions need about $300\,\mathrm{ns}$ to reach the grid. The ions will then induce a tiny signal compared to the electrons, but for a longer time, as described by the Ramo-Shockley theorem Eq.~(\ref{eq:Ramo}). We illustrate the signal induced on the grid in Figure~\ref{fig:sigFormation} using an analytical model that will later be described in Section \ref{sec:Prim}. The peaks represent the electronic signal; the baseline is due to the ionic signal. The electronic noise defines the current detection threshold of the Flash signal. The brown signal of Figure~\ref{fig:sigFormation} illustrates a situation that could happen when operating at low gain: the measured signal stops when the last primary electron enters the gap. The blue signal describes the situation at high gain: the detector gets more sensitive to the ionic contribution and still measures a signal more than $200\,\mathrm{ns}$ after the arrival of the last primary electron. We here stress that, for two identical primary clouds, the measurements differ depending on the gain of the detector.

Being sensitive to the ionic signal even after the arrival of the last primary electron results in an elongation of the measured track and, consequently, in a bias on the reconstruction of the track angle (under-estimation of the polar angle). The interplay between the electronic and the ionic signals could also distort the measurements. However, when controlled, the ionic contribution opens the window for low-energy detection since it acts as a magnification of the signal and it embeds physical information that can be used for directionality as we will later see. We mention that the sensitivity to the ionic signal at high gain does not alter the measurement of the ionization energy of a particle. 
	
	\subsection{The Comimac beamline}

	We briefly introduce Comimac \cite{Comimac} which is a table-top accelerator that can send electrons or ions of controlled kinetic energy inside a MIMAC chamber in the energy range $[150\,\mathrm{eV},~30\,\mathrm{keV}]$. Comimac uses an Electron Cyclotron Resonance Ion Source (ECRIS) that produces a plasma in its resonant cavity using a low-power microwave ($5~\rm{W} - 2.45~\mathrm{GHz}$). A voltage $V_{\mathrm{ex}}$ is applied to extract either electrons or ions (depending on the polarity) with kinetic energy given by $E_K = qV_{\mathrm{ex}}$.
	
	\begin{figure*}
	\centering
	\includegraphics[width=0.7\linewidth]{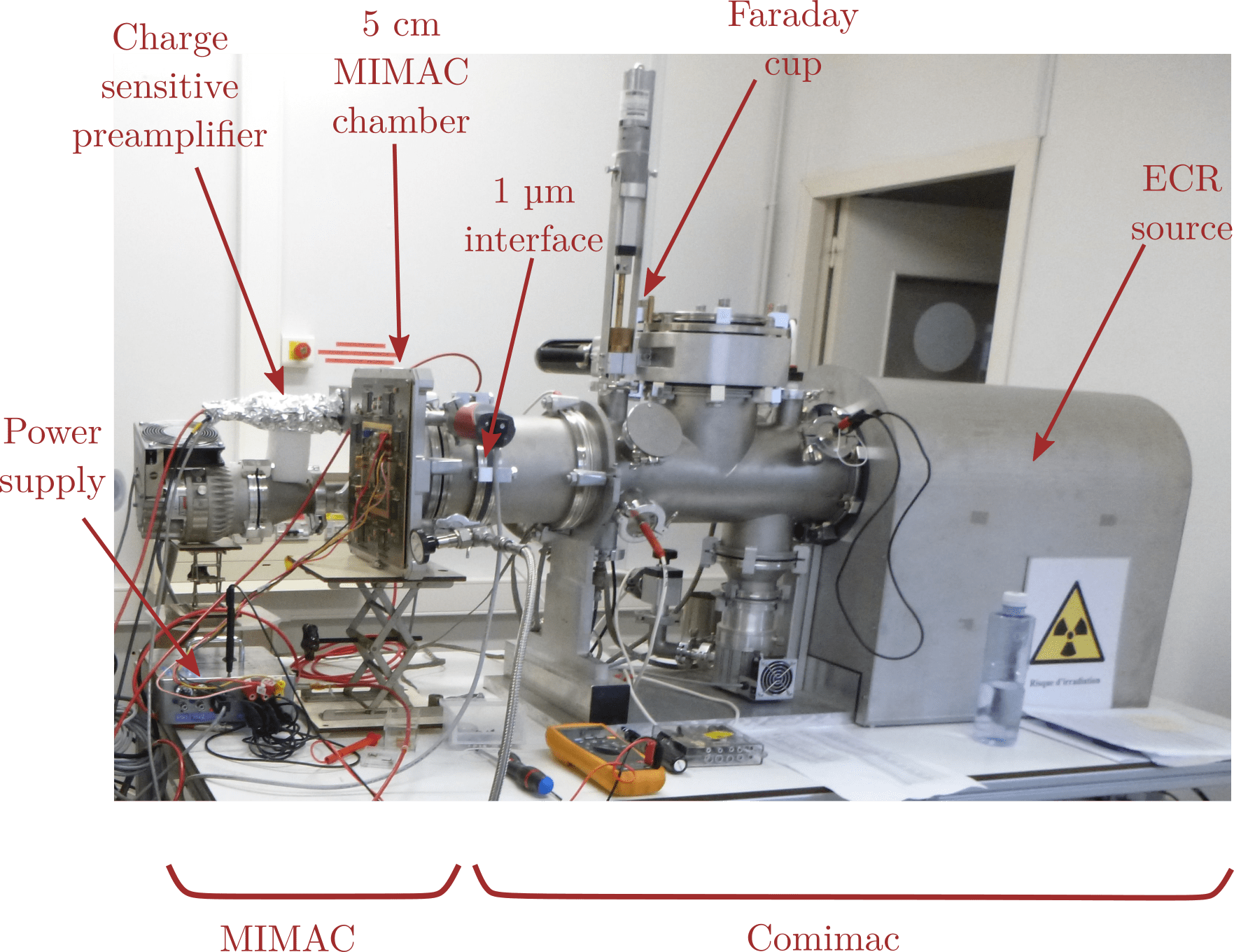}
	\caption{Picture of the experimental setup coupling a MIMAC chamber to the Comimac beamline. The ions and electrons produced by Comimac enter a MIMAC chamber at the cathode level. They are sent parallel to the electric field lines, \textit{i.e.} along the Z-axis of the detector.}
	\label{fig:comimac}
\end{figure*}
	
In order to send the particles into a detector, Comimac is coupled with a MIMAC chamber having $5\,\mathrm{cm}$ of drift. The coupling is performed thanks to a $1\,\mathrm{\mu m}$ hole that ensures pressure independence between the Comimac facility and a MIMAC chamber. A picture of the experimental setup is presented in Figure~\ref{fig:comimac}. The particles are all sent at the same positions in the detector, at the cathode level, with a direction oriented along the electric field lines, \textit{i.e.} along the Z-axis of the detector. In this paper we use Comimac to calibrate the detector with multiple energy points by sending electrons into the chamber. Comimac will also be used to measure the proton IQF in our gas mixture or to test and validate our model as in \ref{sec:SimuMimac} and \ref{sec:Prim}.


\section{Numerical modelling with \texttt{SimuMimac}}\label{sec:SimuMimac}

\subsection{Exclusion of the space-charge effect hypothesis}

The measurements that we have previously published, as explained above, presented a puzzling elongation of the duration of the signal in correlation with the applied voltages. Our first guess to explain this phenomenon was to consider a space-charge built by the large number of ions accumulating in the gap, in particular close to the grid wires due to the topology of the electric field lines. We supposed that the Coulomb field of the accumulated ions could be sufficient to influence the motion of the primary charges and the avalanches development. Since our measurements were not dependent on the event rate, we made the hypothesis of a space-charge built within a single event: the last primary electrons to arrive at the grid would be influenced by the ions produced during previous avalanches. In other words, such a space-charge effect would correspond to an event-based self-shielding of the charges in the gap. 

	To the best of our knowledge, there is no available toolkit able to simulate a space-charge effect in a Micromegas. The main reason is that usual simulations track charges one-by-one whereas we need to transport all particles within one timestep and to compute the local distortions of the electromagnetic field before moving to the next timestep. We have then implemented our own simulation code written in C++, \texttt{SimuMimac}, that models the signal induced on the grid and the anode strips of MIMAC in 2D. The user just has to define the primary electrons cloud (obtained from Monte-Carlo or by running \texttt{SRIM} \cite{SRIM} within the code) and the experimental conditions. The code handles the transport of the electrons and the ions, the avalanches, and the signal formation. \texttt{SimuMimac} is inspired by the code of \texttt{Garfield++} \cite{Veenhof1998, Schindler2021} and partially relies on it for the computation of some fields. The local distortions of the electromagnetic (EM) field are computed analytically from the map of all ions using a covariant formalism based on the Liénard-Wiechert potentials. The working principles of \texttt{SimuMimac} are detailed in Appendix \ref{app:SimuMimac}. Note that the weighting field is not sensitive to the presence of a space-charge \cite{DeVisschere1990, Hamel2008}, so the Ramo-Shockley theorem would remain valid in this situation.
	
	As an important result, \texttt{SimuMimac} has not shown any evidence for a space-charge effect, even when operating at a high gain of about $5\times 10^4$ secondary charges per avalanche (obtained in the Mimac gas at $50\,\mathrm{mbar}$ with an amplification field of $11.1\,\mathrm{kV/cm}$) and for a dense cloud of electrons separated by $1\,\mathrm{ns}$ at grid, all of them entering the gap at the same position (worst case scenario). For the same primary electrons cloud, no significant deviation is observed between the \texttt{SimuMimac} output obtained by considering the local distortions of the EM field and the one without considering the distortions. We later present experimental data in Section \ref{eq:deconv} for which no non-linear effects (like space-charge effects) are observed below a gain of $10^5$. This result allowed to deeply fasten the code by modifying its implementation: we no longer compute the distortions of the EM field. With such changes, the computing times of \texttt{SimuMimac} and \texttt{Garfield++} are similar. The code of \texttt{SimuMimac} is available on request to the authors.
	
	\subsection{Agreement between simulations and measurements}
	
The main success of \texttt{SimuMimac} has been to reproduce the measurements published in \cite{Tao2019} for which we have sent fluorine ions at the cathode with kinetic energies in the range $[6.3\,,\, 26.3\,\mathrm{keV}]$ by the use of a facility similar to Comimac. For this experiment, a MIMAC chamber with $5\,\mathrm{cm}$ of drift is filled with $50\,\mathrm{mbar}$ of \textit{Mimac gas}, \textit{i.e.} a mixture of 70\% CF$_4$ + 28\% CHF$_3$ + 2\% i-C$_4$H$_{10}$. We use an amplification field of $11\,\mathrm{kV/cm}$ with an avalanche gain measured around $2.2\times 10^4$ using a W-value of $38\,\mathrm{eV}$ (mean energy required to form an electron-ion pair).

In order to reduce the statistical fluctuations due to diffusion, we have decided to separate the simulations in two steps: we first transport the primary charges in the drift region with \texttt{Garfield++}; secondly, we simulate the avalanches and compute the induced signals with \texttt{SimuMimac}. In the drift region, we transport a large number of charges ($>3\times10^4$) in order to reduce uncertainties inherent to stochastic processes, such as diffusion for instance. For this simulation, the primary electrons cloud is generated with \texttt{SRIM}. The second step of the simulation uses the results of the first step as input: the coordinates of each primary electron at the grid level are drawn from a 2D Gaussian distribution whose properties have been defined thanks to the results of the first step. \texttt{SimuMimac} is finally called to simulate the avalanche of the electrons and to determine the signals induced on the sensors. This "two-step simulation" enables us to operate with low statistics for the \texttt{SimuMimac} simulation. The statistics depend on the computing time and vary between 20 simulated fluorine ions (at $26.3\,\mathrm{keV}$) and 150 (at $6.3\,\mathrm{keV}$). 

\begin{figure*}
	\centering
	\includegraphics[width=0.95\linewidth]{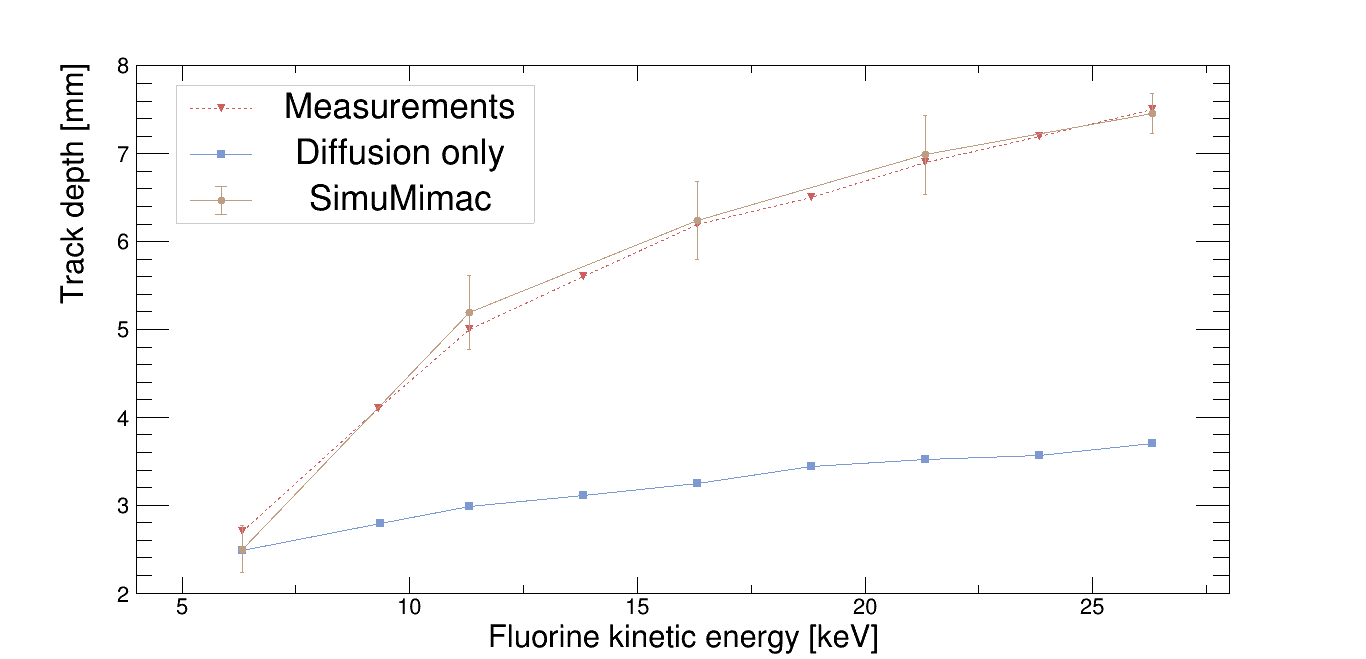}
	\caption{Fluorine track depths (Z-component of the track) after $5\,\mathrm{cm}$ of drift in the experimental conditions of \cite{Tao2019}. Measurements are shown in red; the blue squares refer to a simulation that does not take into account the signal induced by the ions in the Micromegas gap; the brown dots are obtained with \texttt{SimuMimac} where the error bars account for the statistical uncertainty at one sigma.}
	\label{fig:LHIsimuMimac}
\end{figure*}

As a drawback of \texttt{SimuMimac}, we must define manually two parameters of the simulations. One must adjust the mean mobility of the ions in the gap based on an extrapolation from database values, more details can be found in Appendix \ref{app:SimuMimac}. The second parameter to fix is the strip threshold on the pixelated anode, \textit{i.e.} the current above which the strips are fired. A mean ion velocity of $1.7\,\mathrm{\mu m/ns}$ and a strip threshold of $0.56\,\mathrm{\mu A}$ led to the best agreement with measurements. The values are adjusted manually for one energy that is used as reference, and we later compare the tendencies of the simulated quantities to the measurements in order to quantify their agreement. 

We here discuss measurements and simulations of track depths, that is the Z-component of the track. The results are presented in Figure~\ref{fig:LHIsimuMimac} where the only novelty compared to our previous work \cite{Tao2019} is the \texttt{SimuMimac} simulation. The blue data points correspond to the \texttt{Garfield++} simulation described above that does not take the avalanche into account. \texttt{SimuMimac} agrees with the measured data within the statistical uncertainty. The measured depth appears as elongated (compared with the expected size of the primary electrons cloud before the avalanche) because of the contribution of the ionic signal. The large number of ions accumulated in the gap after the arrival of the last primary electron keeps inducing a detectable signal on the anode strips for several timeslices. This signal depends on the total number of ions located in the amplification area at the same time. In other words, it depends on the kinematics balance in the gap between the input charges (primary electrons starting an avalanche) and the output (ions being collected on the grid). For this reason, the influence of the ionic contribution increases with the charge density of the primary electron cloud, and thus with the kinetic energy of the fluorine ions. One can see in Figure~\ref{fig:LHIsimuMimac} that the ionic signal represents more than half of the total duration of the induced signal for kinetic energies above $16.32\,\mathrm{keV}$.

In our previous work, we proposed an empirical correction of the depth by multiplying the measurements with the \textit{asymmetric factor}. This factor is defined as the relative position of the inflexion point of the Flash signal. In other words, it is the ratio between the duration before reaching the inflexion over the duration after the inflexion point, more details can be founded in \cite{Tao2019}. An asymmetric factor of 1 means that the inflexion point is centred in time. At high gain with a large gap of $512\,\mathrm{\mu m}$, we however measure asymmetric factors between $0.6$ and $0.8$, showing an asymmetry in the time distribution of the signal. \texttt{SimuMimac} can retrieve the asymmetric factor and it agrees with the measurements within $2\%$. We are now able to explain the main origin of this asymmetry: the more the detector is sensitive to the ionic signal, the smaller the asymmetric factor. 

We have developed a simulation code that reproduces the evolution of the track length measurements with respect to the fluorine kinetic energy and that agrees with measured asymmetric factors. \texttt{SimuMimac} brought us to some important conclusions in the understanding of the detector at high gain. First, we are not affected by an event-based space-charge effect in the experimental conditions considered. Second, the signal induced by the ions can alone explain the measured track elongations and the asymmetry of the Flash derivative. Third, the ionic contribution can represent more than half of the measured duration of the signal and must then be precisely studied. We now have a simulation tool giving consistent trends with experiments that we can use to investigate the behaviour of the detector.


\section{Accessing experimentally the primary electrons cloud}\label{sec:Prim}

The physical observables required for directional WIMP (or neutron) searches are retrieved from the properties of the primary electrons generated by the WIMP-induced nuclear recoil: the total number of primary electrons and the coordinates $(x,y,t)$ of each of them. The coordinates are affected by diffusion all along the drift of electrons towards the amplification area. We have seen in previous sections that another process alters the measurements due to the influence of the ions in the amplification region. In other words, the measured signals are distorted both by diffusion and by the ions produced in avalanches. While diffusion slightly blurs the signal, it also plays a key role in directional detection of sub-millimeter tracks (\textit{e.g.} recoils in the keV-range) since it Gaussianly enlarges tracks that would otherwise be detected as point-like events \cite{Tao2019}. The present work focuses on the second phenomenon: the influence of the ions. In this section, we describe how to deconvolve the ionic contribution from the measurements in order to extract the electronic current induced on the grid. The electrons produced in an avalanche, that is initiated by a single primary electron entering the amplification region, are collected in less than $1\,\mathrm{ns}$. This value being significantly lower than the detector time resolution of $20\,\mathrm{ns}$, the electronic current induced on the grid can be seen as the time distribution of the primary electrons cloud before the avalanche. Its integral is proportional to the number of primary electrons.

\begin{figure*}
	\centering
	\includegraphics[width=0.95\linewidth]{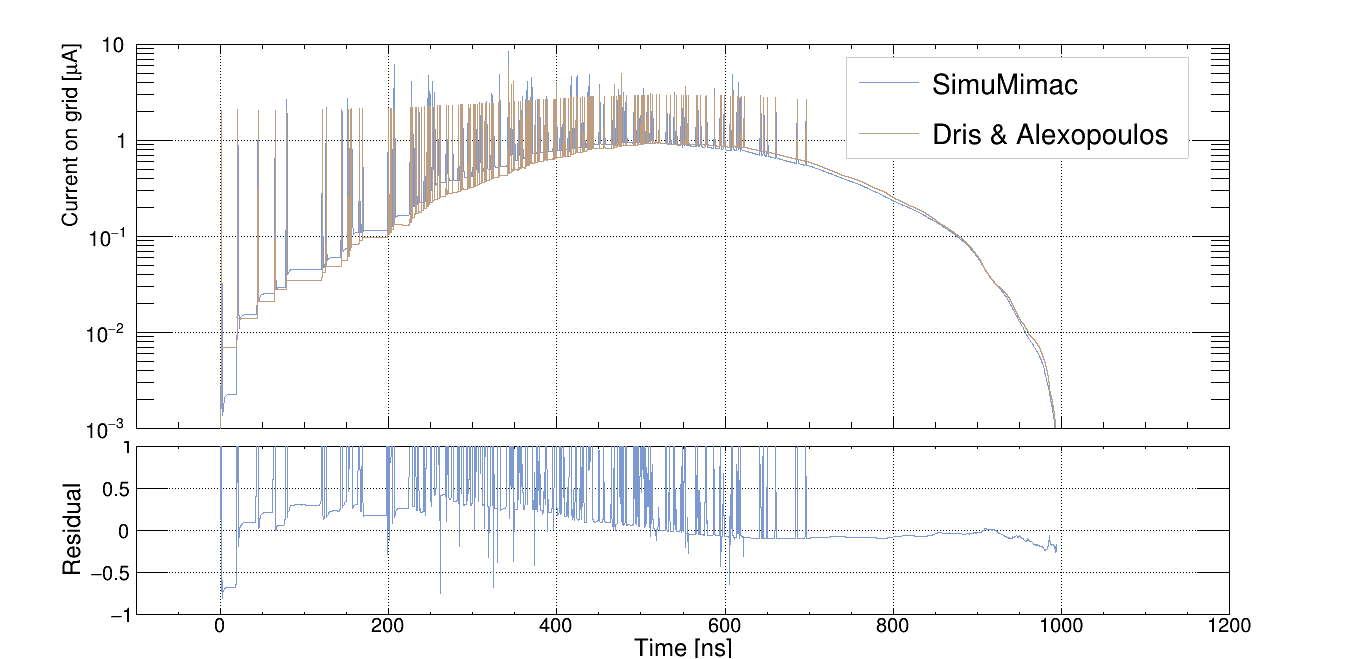}
	\caption{Comparison between \texttt{SimuMimac} and the analytical model from Dris and Alexopoulos. Example of the current induced on the grid for a $10\,\mathrm{keV}$ protons after $5\,\mathrm{cm}$ of drift in a gas mixture of 50\% i-C$_4$H$_{10}$ + 50\% CHF$_3$ at $30\,\mathrm{mbar}$. The lower panel presents the residual $R = (S-M)/M$ where $S$ and $M$ are the \texttt{SimuMimac} and the analytical model signals, respectively.}
	\label{fig:DrisSimuMimac}
\end{figure*}

\subsection{Deconvolution of the ionic contribution}

Dris and Alexopoulos \cite{Dris2014} have estimated analytically the signal formed on the grid of a Micromegas by a single avalanche:
\begin{equation}
\begin{cases}
	f(t) &= \frac{qu_n}{d}\,e^{\alpha u_nt} \hspace*{1.75cm}\mathrm{,~for~} 0 \leq t \leq d/u_n\\
	g(t) &=\frac{qu_p}{d}\bigg(e^{\alpha d} - e^{\alpha u_pt}\bigg)\hspace*{0.3cm}\mathrm{,~for~} 0 \leq t \leq d/u_p
	\label{eq:Dris}
\end{cases}
\end{equation}
where $f(t)$ is the electronic current induced on the grid, $g(t)$ is the ionic one, $q$ is the elementary charge, $u_n$ is the mean electron drift velocity in the gap, $u_p$ is the mean ion drift velocity in the gap, $d$ is the gap length, and $\alpha$ is the Townsend coefficient. This elegantly simple formula relies on several approximations. The main ones are: one-dimensional calculations, a constant Townsend coefficient, a mean ion mass produced in the avalanche, the grid and the anode are considered as having geometry of an ideal plane capacitor. A comparison between the Dris and Alexopoulos model and \texttt{SimuMimac} is presented in Figure~\ref{fig:DrisSimuMimac} for a typical example of a $10~\rm{keV}$ proton after $5~\rm{cm}$ of drift. While some differences are observed, mainly on the amplitude of the electronic peaks, the overall agreement between the two methods is better than $50\%$, which qualitatively validates the use of the Dris and Alexopoulos model for determining the signal induced on the Micromegas grid. We can then make use of two complementary approaches to determine the signal induced on the grid: a MC simulation and an analytical model. Note that the signal induced on the anode is more difficult to model than on the grid, since the non-linear weighting fields near the anode strips lead to complex analytical expressions in the Dris and Alexopoulos model. For this reason, we do not provide an analytical model of the signal induced on the anode strips, and we instead use \texttt{SimuMimac} to simulate it when required.

As explained before, we aim to access the electronic current induced on the grid. The Flash signal measured by MIMAC corresponds to the integrated charge induced on the grid, which results from a convolution between the electronic and the ionic contributions. We can analytically deconvolve the ionic contribution in order to retrieve the electronic current. The details of the calculations can be found in Appendix \ref{app:deconvolution}. The electronic current is proportional to the following expression:
\begin{equation}
		f(t_i) ~\propto~ D(t_i) - D(t_{i-1})  - e^{-A\Delta t}\Big\lbrace D(t_{i-1}) - D(t_{i-2})\Big\rbrace
		\label{eq:deconv}
\end{equation}
where $D(t_i)$ is the derivative of the Flash signal at the timeslice $t_i$. The parameter $A$ can directly be estimated from the Flash signal, as explained in Appendix \ref{app:deconvolution}.  It means that the electronic current can be retrieved from any MIMAC measurements using any gas mixture, with no use of prior nor introduction of \textit{ad hoc} parameters. To be mentioned, this expression suffers from two drawbacks: (1) the sampling time of $20\,\mathrm{ns}$ is large compared with the time separation between two primary electrons, which limits the resolution of the deconvolution; (2) in the derivation we use an approximation assuming that the primary electrons cloud at grid is uniform. We also mention that the deconvolution must be applied to data after the correction of the ballistic deficit introduced by the time response of the charge sensitive preamplifier. The correction is performed following the method described in \cite{Gavrilyuk2015} with a measured time constant of $16\,\mathrm{\mu s}$.

\begin{figure*}
	\centering
	\includegraphics[width=0.95\linewidth]{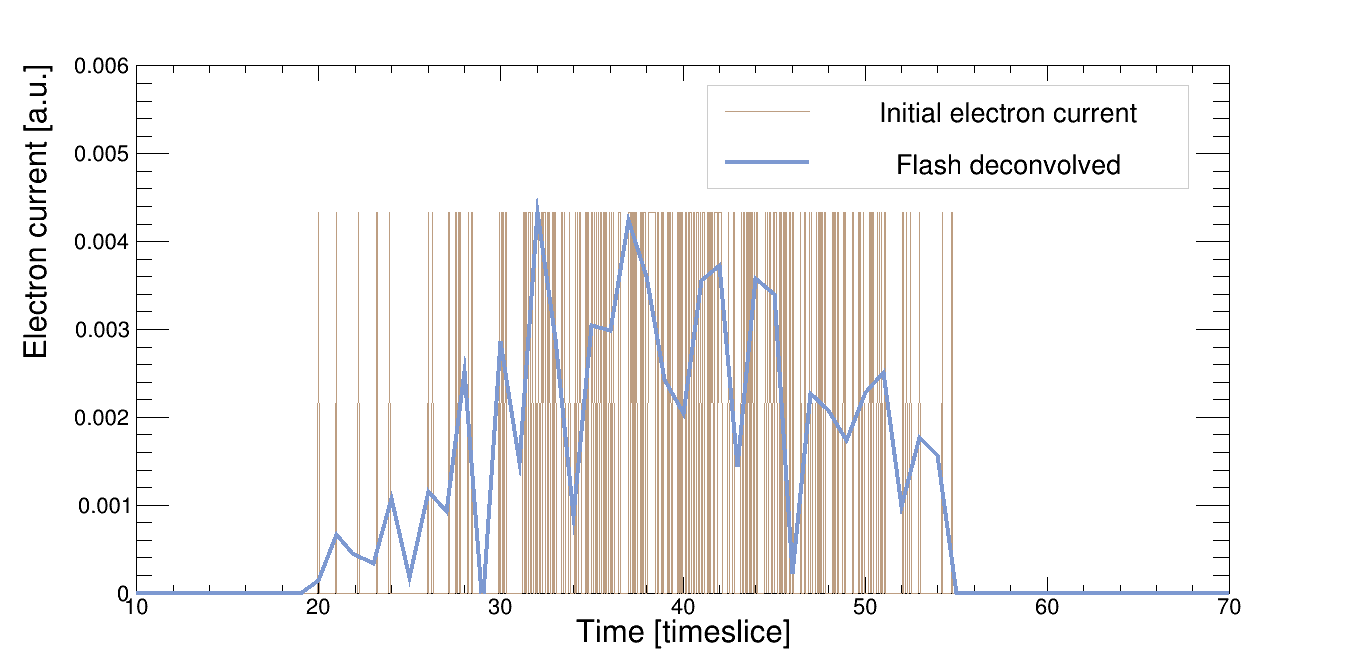}
	\caption{Deconvolution of the \texttt{SimuMimac} simulation of the $10\,\mathrm{keV}$ proton presented in Figure~\ref{fig:DrisSimuMimac}. The time distribution of the electron cloud at the grid is shown in brown. The blue curve is obtained by applying Eq.~(\ref{eq:deconv}) on the simulated Flash signal which has been discretized in MIMAC-like timeslices of $20\,\mathrm{ns}$.}
	\label{fig:deconvolution}
\end{figure*}

\subsubsection*{Evaluation of the deconvolution on simulated data}

The performances of the deconvolution can be determined on simulated data. One can qualitatively appreciate in Figure~\ref{fig:deconvolution} the deconvolution of a $10\,\mathrm{keV}$ proton simulated with \texttt{SimuMimac}: the electron current obtained by deconvolution follows the time distribution of the initial cloud. One can also quantify the bias on the time of arrival of the last primary electron by comparing the initial cloud with the deconvolved Flash signal. To do so, we have implemented a simple Monte-Carlo code that generates $n$ clouds with properties drawn randomly (number of electrons, length, polar angle) and we transport them to the grid by Gaussianly applying diffusion coefficients obtained with Magboltz. We then use the model from Dris and Alexopoulos, Eq.~(\ref{eq:Dris}), to determine the signal formed on the grid. Finally, the last step consists in deconvolving the signal with Eq.~(\ref{eq:deconv}). We obtain a bias of $-0.6 \pm 1.6\,\mathrm{timeslice}$, meaning that the deconvolution slightly under-estimates the time of arrival of the last primary electron. This bias remains small compared to the duration of the electronic signal ($\sim35\,\mathrm{timeslices}$ in the considered case).

\subsubsection*{Evaluation of the deconvolution on experimental data}

\begin{figure*}
	\centering
	\includegraphics[width=0.95\linewidth]{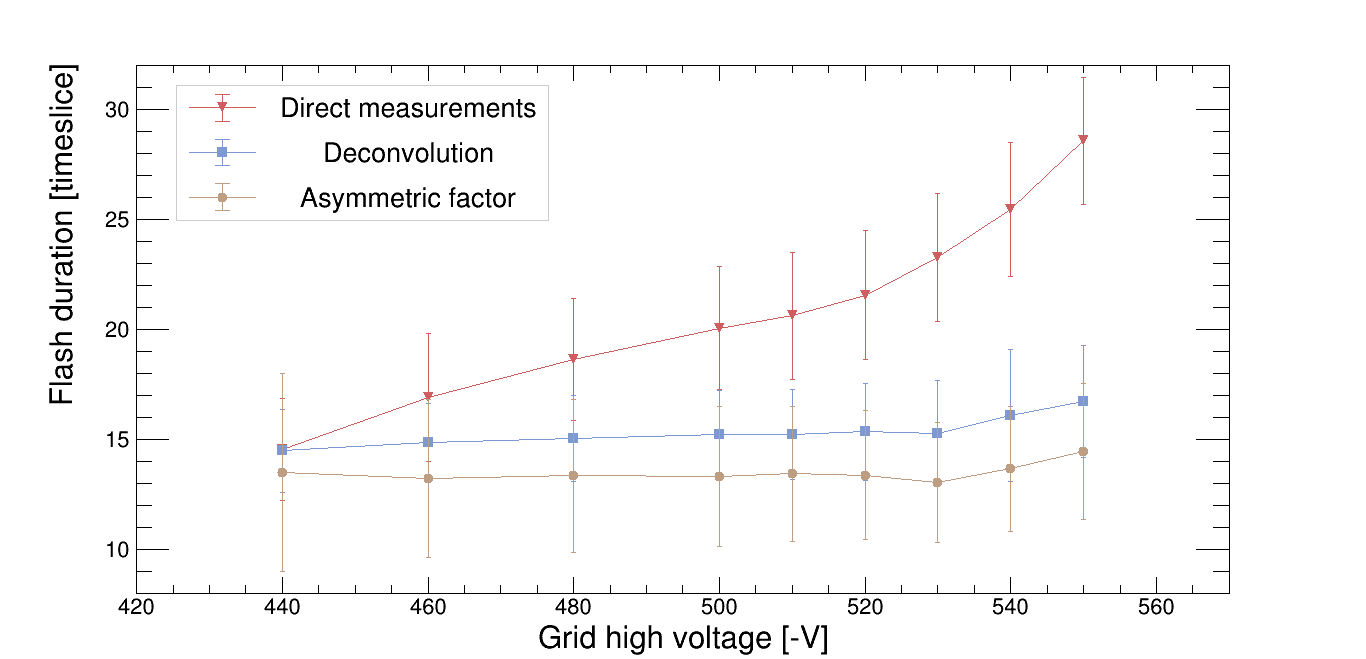}
	\caption{Gain curve measured by varying the high-voltage applied on the grid. For each gain, Comimac sends $5\,\mathrm{keV}$ electrons in a gas mixture of 50\% i-C$_4$H$_{10}$ + 50\% CHF$_3$ at $30\,\mathrm{mbar}$. The duration of the Flash is shown in red. The blue curve represents the duration after deconvolution of the ionic contribution. The brown curve is obtained by multiplying the Flash duration with the asymmetric factor described in Section \ref{sec:SimuMimac}. The error bars represent the statistical uncertainties at one sigma.}
	\label{fig:gainCurve}
\end{figure*}

We can also evaluate the performances of the deconvolution on experimental data. We use Comimac to send $5\,\mathrm{keV}$ electrons in a gas mixture of 50\% i-C$_4$H$_{10}$ + 50\% CHF$_3$ at $30\,\mathrm{mbar}$ and we vary the high voltage applied on the grid. We know that when the gain increases, the detector gets more sensitive to the ionic contribution, so we measure a longer Flash signal. However, the deconvolution gives access to the time distribution of the primary electrons cloud, so it must not depend on the gain. The results are presented in Figure~\ref{fig:gainCurve}: the direct measurements show a non-linear trajectory increasing with the gain while the duration of the deconvolved Flash follows a plateau (down to $-530\,\mathrm{V}$), corresponding to a gain-independent duration as expected. The Flash duration is determined from the local minima of the derivative of the Flash signal, illustrated with the blue triangles in Figure~\ref{fig:recoilExample}. For completeness, we also present in Figure~\ref{fig:gainCurve} the effect of the empirical correction based on the asymmetric factor from our previous work: it follows similar tendencies than the deconvolved Flash but has an offset due to the fact that the asymmetric factor integrates both the electronic and the ionic contributions. 

In Figure~\ref{fig:gainCurve}, one can observe a change of regime for high voltages below $-530\,\mathrm{V}$. According to \texttt{SimuMimac}, the gain in such conditions exceeds $10^5$. We propose the hypothesis that at such a very high gain a space-charge starts to affect charge collection, which could distort the signal as explained in Section \ref{sec:SimuMimac}. The test of this hypothesis is out of the scope of the current work since we here present experimental measurements obtained in the plateau region of the gain curve of Figure~\ref{fig:gainCurve}. However, up to a gain of $10^5$, the deconvolution of the Flash signal leads to a constant measured duration that does not vary with the gain. This is an experimental validation of the deconvolution of the ionic contribution.  

\begin{figure*}
	\centering
	\includegraphics[width=0.95\linewidth]{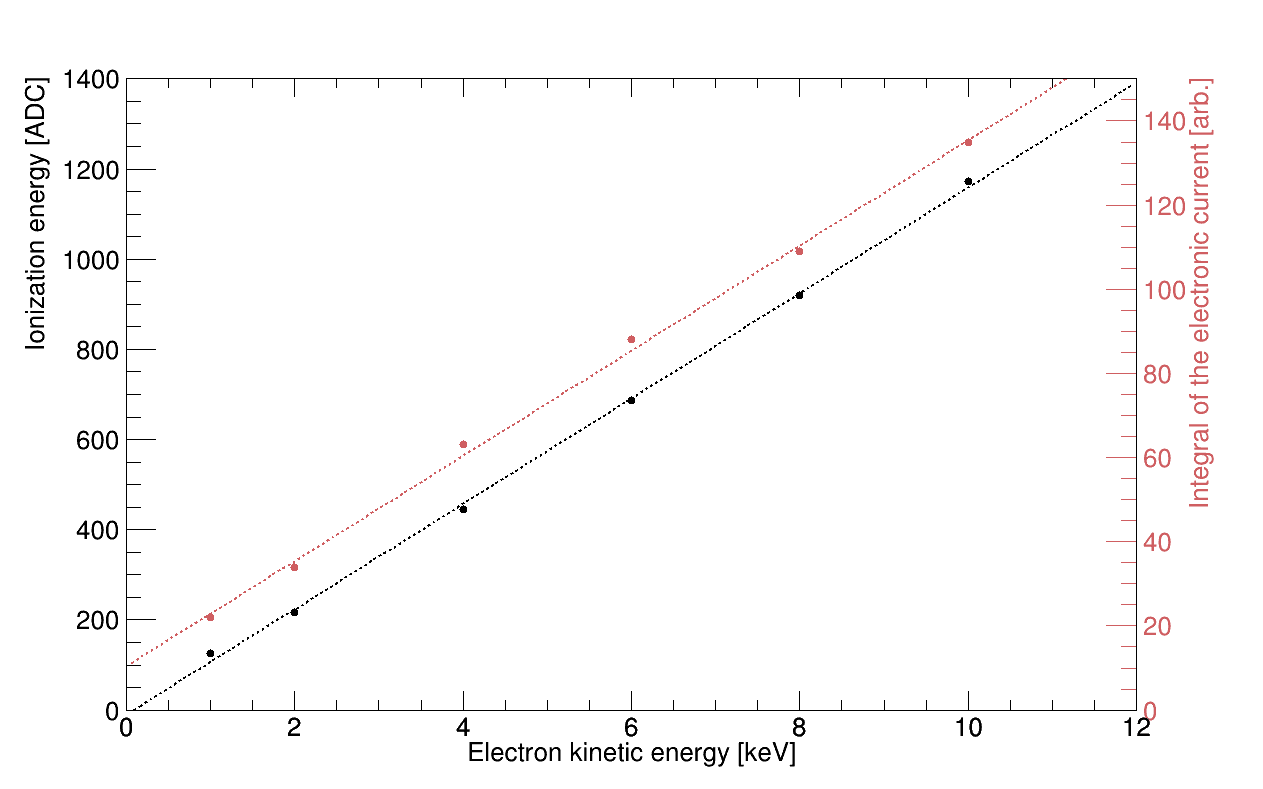}
	\caption{Electron calibration in the \textit{Mimac gas} at $50\,\mathrm{mbar}$ with a gain around $2\times 10^4$. The kinetic energy of the electrons is given by Comimac, the ionization energy (black dots) is derived from the amplitude of the Flash signal, and the red data points represent the integral of the electronic current obtained by deconvolution.}
	\label{fig:calibrationMimacGas}
\end{figure*}

Finally, the integral of the electronic current (resulting from the deconvolution of the ionic signal) should be proportional to the number of primary electrons, thus it should be linear with the kinetic energy. An electron calibration performed with Comimac in the \textit{Mimac gas} at $50\,\mathrm{mbar}$ is presented in Figure~\ref{fig:calibrationMimacGas}. As expected, the integral of the electronic current obtained by deconvolution follows the same tendency as the ionization energy and consequently demonstrates that the deconvolution does not affect the number of primary electrons. However we observe the presence of an offset: a null kinetic energy will still produce a non-zero electronic current. This offset is due to the approximations performed in the derivation of Eq.~(\ref{eq:deconv}) and also in the determination of the parameter $A$, \textit{c.f.} Appendix \ref{app:deconvolution}.

\subsection{Head-tail recognition}

In addition to the measurements of the direction of a nuclear recoil, directional detectors aim to distinguish between the \textit{head} and the \textit{tail} of a track (we define the tail as the closest point to the initial collision). Head-tail recognition plays a crucial role in the discrimination of the background \cite{Billard2011, OHare2015} and it reduces by about one order of magnitude the number of events required for directional detection \cite{Burgos2008}. The usual approach for head-tail recognition relies on the detection of an asymmetry in the charge distribution of the measured signal, the asymmetry being correlated to the stopping power of the nuclear recoil as a function of its kinetic energy. We have seen in Section \ref{sec:SimuMimac} that the ionic signal induces a similar asymmetry at high gain in a large gap of $512\,\mathrm{\mu m}$. For this reason, it is mandatory to separate the asymmetry due to the ionic contribution, from the asymmetry due to the stopping power. 

We have experimentally validated that the deconvolution of the Flash signal gives access to the time distribution of the primary electrons cloud at the Micromegas grid, its integral being linearly correlated with the ionization energy. This approach allows to better describe the fine structure of the primary electrons cloud since the kinematics of the electronic current is hundreds times faster than the one of the ionic current. For the kinetic energies and the gas conditions considered in this work, the stopping power ($-dE/dx$) of a proton decreases when its kinetic energy decreases \cite{SRIM}. In other words, the Bragg peak is located at the beginning of the track and more charges are deposited close to the tail. This signature based on the stopping power can be used to distinguish between the head and the tail of the track.

\begin{figure*}
	\centering
	\begin{minipage}{0.49\linewidth}
	\includegraphics[width=\linewidth]{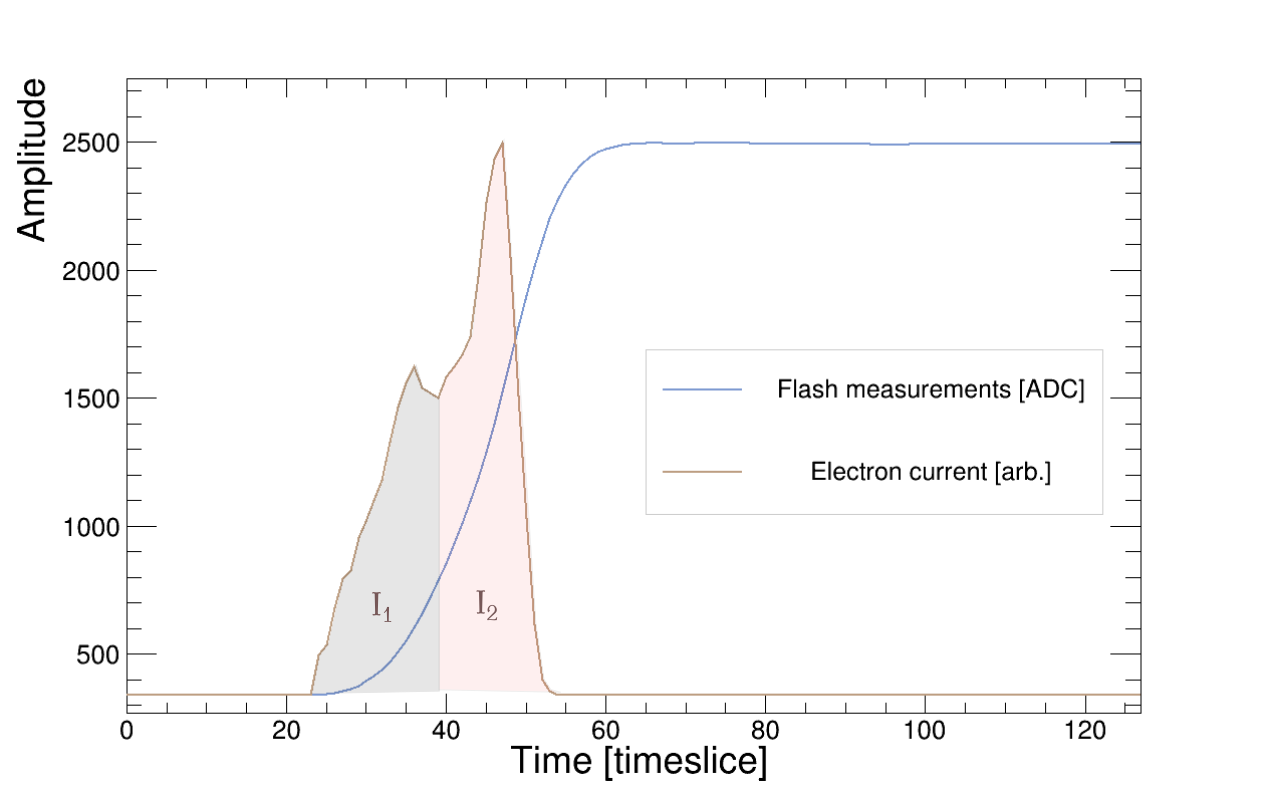}
	\subcaption{Example of a measured event. The electron current has been scaled to appear as high as the Flash signal.}
	\label{fig:deconvExample_13keV}
	\end{minipage}
	\hfill
	\begin{minipage}{0.49\linewidth}
		\includegraphics[width=\linewidth]{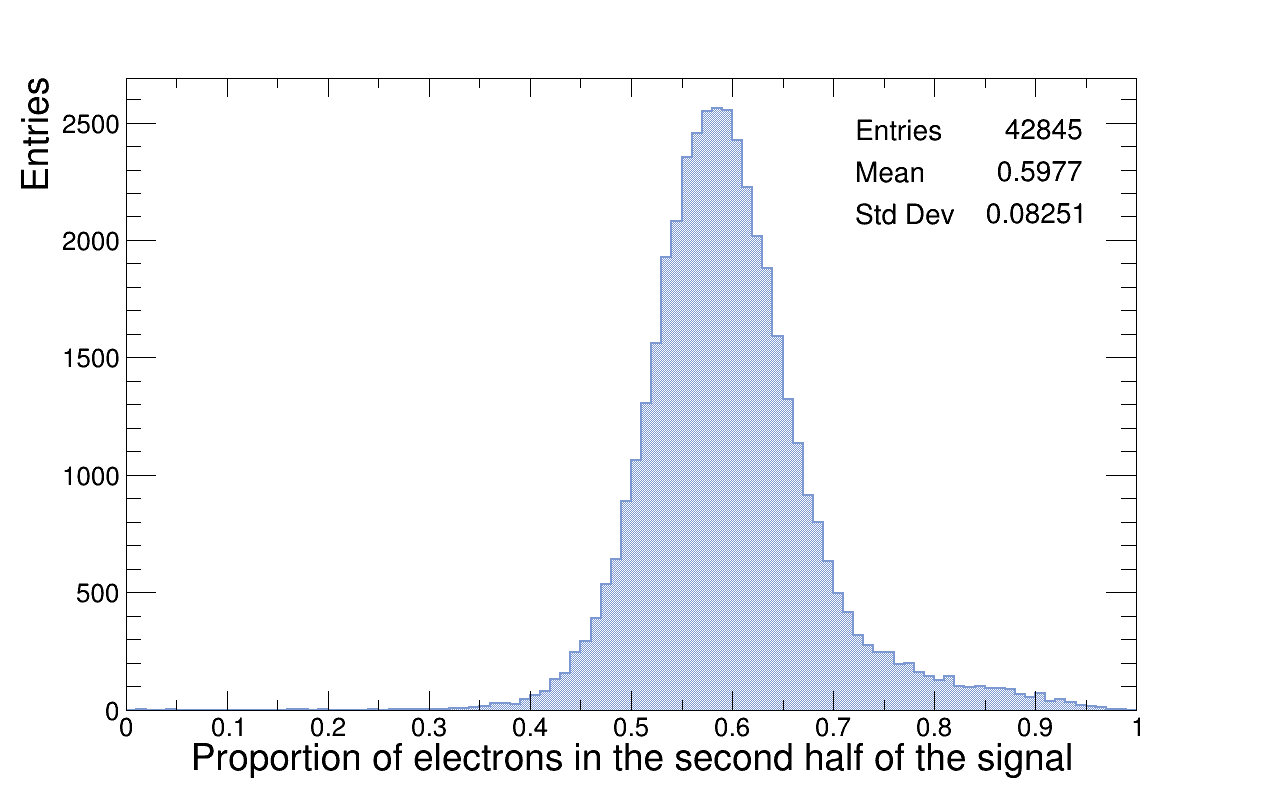}
		\subcaption{Histogram of the proportion of electron current located in the second half of the signal for the entire run.}
		\label{fig:proportionElec}
	\end{minipage}	
	\caption{The asymmetry of the time distribution of the electronic current measured for $13~\rm{keV}$ protons sent by Comimac in a mixture of 50\% i-C$_4$H$_{10}$ + 50\% CHF$_3$ at $30\,\mathrm{mbar}$. In the left panel, the filled areas correspond to the integrals, $I_1$ and $I_2$, of the electron current in the first and the second half of the signal, respectively. In the right panel, the proportion plotted is defined as $I_2/(I_1+I_2)$. In this situation, since the ions are sent at the cathode, the last timeslices correspond to the first interactions of the track (\textit{i.e.} to the tail).}
		\label{fig:headTail}
\end{figure*}

In Figure~\ref{fig:deconvExample_13keV} we present a typical example of the measurements of a $13\,\mathrm{keV}$  proton sent by Comimac in a mixture of 50\% i-C$_4$H$_{10}$ + 50\% CHF$_3$ at $30\,\mathrm{mbar}$. The brown curve is the electronic current obtained by deconvolution of the Flash signal. We have normalized it to appear as high as the Flash. One can see an asymmetry in the time distribution of the electronic current: most of the primary electrons arrive in the second half of the signal. In a Comimac experiment the ions are sent at the cathode, so the last timeslices correspond to the first interactions of the track. The asymmetry presented in Figure~\ref{fig:deconvExample_13keV} follows the expected tendency from the stopping power of the protons.

It is possible to quantify this asymmetry by measuring the integral of the electronic current in the first and second half of the signal, called $I_1$ and $I_2$ respectively in the figure. The ratio $R = I_2/(I_1+I_2)$ describes the proportion of the electronic current located in the second half of the signal. This ratio determines if the track is oriented along $+\vec{\hat{z}}$ or $-\vec{\hat{z}}$. When $R > 0.5$, the tail is located at the end of the signal so the track is oriented towards the grid; and towards the cathode for $R < 0.5$. In the example of Figure~\ref{fig:deconvExample_13keV} we obtain $R=0.61$. Figure~\ref{fig:proportionElec} shows the distribution of $R$ over an entire run. Its mean value at $R=0.598$ indicates that the tracks are mainly oriented towards the grid, as expected from the experiment geometry. Note that in these measurements,  the proton ionization energy is well defined with an energy resolution of 17\% (the FWHM divided by the mean value). We can conclude that the deconvolution of the ionic signal leads to head-tail recognition, event-by-event, since it reveals the asymmetry in the time distribution of the primary electrons cloud.


\section{Directionality on nuclear recoils in the keV-range}\label{sec:directionality}

We illustrate in this section how the deconvolution of the ionic signal can lead to additional observables for directionality. This work is exploratory and aims to open windows for improving usual methods (based on 3D track reconstruction) involved in directional detections. The directional performances of the MIMAC detector can be evaluated experimentally in a mono-energetic neutron field. The elastic scattering of a neutron on a nucleus of our gas mixture will induce a nuclear recoil. We will here only consider proton recoils, since in a mixture of 50\% i-C$_4$H$_{10}$ + 50\% CHF$_3$ they represent $87\%$ of the total nuclear recoils, according to stoichiometry and cross-sections from the \texttt{ENDF} database \cite{ENDF}. The kinetic energy of proton recoil induced by an elastic collision with a neutron, $E_p$, can be expressed in the lab frame as:
\begin{equation}
	E_p = E_n~\cos^2\theta
	\label{eq:recoilE}
\end{equation}
where $E_n$ is the neutron kinetic energy and $\theta$ the scattering angle (between the incident neutron and the proton recoil directions) with an angular distribution centred on $45\,^\circ$. For a mono-energetic neutron field, the simultaneous measurement of $E_p$ and $\theta$ allows to reconstruct the energy spectrum. One can then evaluate the directional performances of a detector by comparing the reconstructed neutron energy spectrum with the expected one. 

As introduced in Section \ref{sec:sigForm}, the Ionization Quenching Factor (IQF) of protons in our gas mixture must be determined in order to convert the measured ionization energy into the kinetic energy, $E_p$, before applying Eq.~(\ref{eq:recoilE}). The IQF can be simulated by \texttt{SRIM} from an extension of the Lindhard theory \cite{Lindhard} coupled with experimental data. However, we have previously shown large discrepancies between IQF measurements and simulations below $50\,\mathrm{keV}$ \cite{Guillaudin2012, Santos2008} in low-pressure gas mixtures. For this reason, we have performed a measurement of the proton IQF in our gas mixture, following the procedure described in \cite{Beaufort2022}. The direct comparison of the ionization energy of protons and electrons (of the same kinetic energy) sent by Comimac gives access to the protons IQF. Following the parametrization of \cite{Mei2008}, we obtain $\mathrm{IQF}(E_p) = E_p^\alpha/(\beta+E_p^\alpha)$ with $\alpha = 0.24$ and $\beta = 1.20$. Our measurements follow similar tendency than \texttt{SRIM}'s IQF but shifted to lower IQF by about $38\%$.

	\subsection{Experimental setup}\label{subsec:expNeutron}
	
\begin{figure*}
	\centering
	\includegraphics[width=0.6\linewidth]{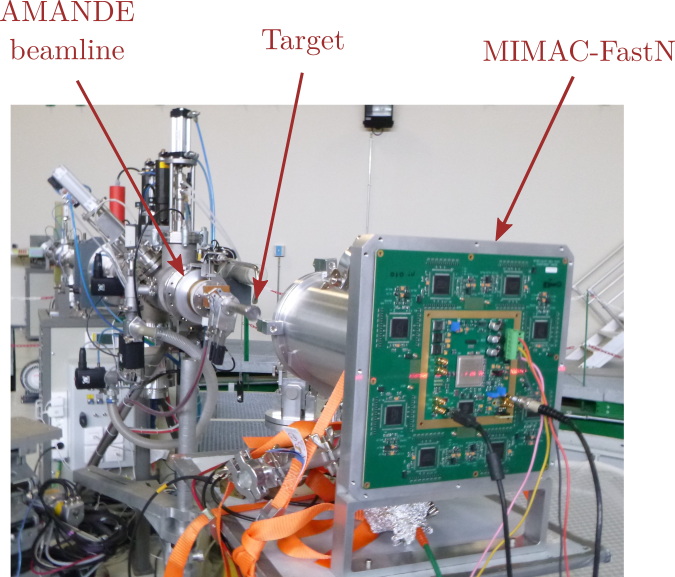}
	\caption{Experimental setup of the AMANDE campaign.}
	\label{fig:amande}
\end{figure*}

The AMANDE facility \cite{Gressier2008, Gressier2014} of the French Institute for Radiation protection and Nuclear Safety (IRSN) produces mono-energetic neutron field between $2\,\mathrm{keV}$ and $20\,\mathrm{MeV}$. In this section, the AMANDE facility makes use of the nuclear reaction $^{45}\mathrm{Sc}\,(p,n)$ to produce neutron fields with kinetic energy of $8.12\pm0.01\,\mathrm{keV}$ or $27.24\pm0.05\,\mathrm{keV}$ \cite{Maire2016} depending on the energy of the proton beam that activates one resonance or another. A MIMAC chamber specially designed for neutron spectroscopy, MIMAC-FastN \cite{Sauzet2019}, is placed in front of the $^{45}\mathrm{Sc}$ target, at a distance of $33\,\mathrm{cm}$, such that the proton beam is parallel to the Z-axis of the detector. A picture of the experimental setup is shown in Figure~\ref{fig:amande}. The chamber is filled with a gas mixture of 50\% i-C$_4$H$_{10}$ + 50\% CHF$_3$ at $30\,\mathrm{mbar}$ stored in a buffer volume that has previously been used to perform the energy calibration of the detector with Comimac by sending electrons of multiple kinetic energies in between $3\,\mathrm{keV}$ and $15\,\mathrm{keV}$.

Photons are also produced during the nuclear reaction on the $^{45}\mathrm{Sc}$ target with a fluence about $20$ times larger than the neutron's one \cite{Maire2016}. The electron-recoil discrimination represents then a central element of our analysis in order to extract the proton recoils from the $\gamma$ background. We first apply a few \textit{minimal cuts}, as detailed in \cite{Riffard2016}, that suppress almost $98\%$ of the events that are easily identified as non-recoil events. After this step, the strategies for electron-recoil discrimination differ between the two datasets presented below: the one at $27\,\mathrm{keV}$ and the one at $8\,\mathrm{keV}$.

At $27\,\mathrm{keV}$, we follow the same procedure as in our previous analyses \cite{Riffard2016} in training a Boosted Decision Tree (BDT). We perform a "background only" run for which the energy of the AMANDE proton beam is sufficiently decreased to remain out of the neutron resonance: the proton beam, only $20\,\mathrm{keV}$ less energetic at the same current level \cite{Maire2016}, keeps producing the same $\gamma$ rate on the target while no neutron gets produced. This "background only" run is combined with the measurements (corresponding to a "signal + background" run) in order to train and test a BDT thanks to the \texttt{TMVA} software \cite{TMVA}. The discriminating observables given to the BDT are mainly derived from the event tracks (\textit{e.g.} number of holes). The background rejection power, that is the number of background events rejected for a single background event passing the cuts, can be determined on the test sample: we obtain $8\times10^4$. The expected number of recoils can be roughly estimated from the event rate difference between the background run and the measurements in the neutron field. The comparison between this expected number and the effective number of recoils after application of the BDT ($2300$ events) gives an estimation of the BDT acceptance: about $50\%$ of the proton recoils are kept. For directional analyses, the exclusion of recoil events does not represent an issue except that it decreases the statistics.

For the dataset with neutrons of $8\,\mathrm{keV}$, we measure proton recoils down to $500\,\mathrm{eV}$. In these conditions the BDT is not uniform: it accepts more recoils at large energy than low energy. This non-uniformity would introduce a bias on the angle reconstruction so we instead decide to implement a standard discrimination based on track observables, as in \cite{Sauzet2019}. As an important drawback, this approach rejects fewer background events than the BDT but it accepts almost all recoil events. We estimate that $30\%$ of the kept events are due to the background by comparing the cut efficiencies at $8\,\mathrm{keV}$ and $27\,\mathrm{keV}$ as well as the neutron production cross-sections \cite{Cosack1985} ($10\%$ larger neutron production at $8~\mathrm{keV}$). For this dataset we keep $5400$ events.
	
	\subsection{Directionality from the deconvolution of the ionic contribution}
	
\begin{figure*}
	\centering
	\begin{minipage}{0.49\linewidth}
		\includegraphics[width=\linewidth]{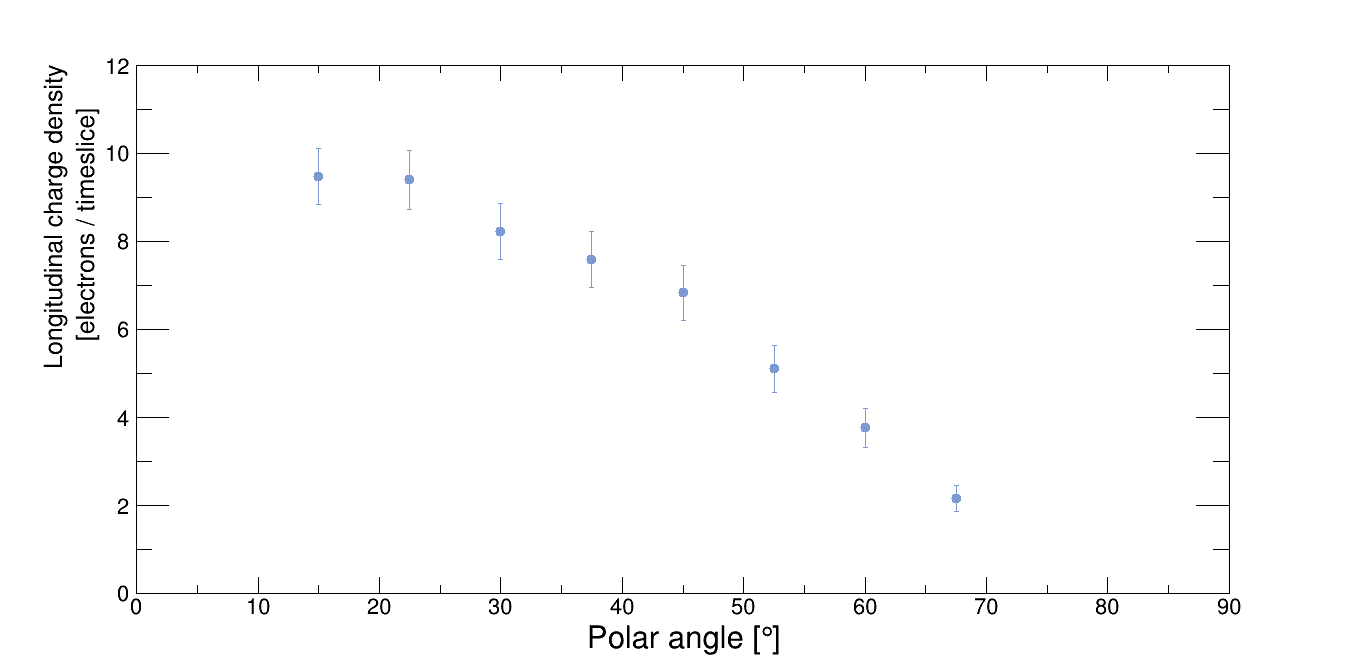}
		\subcaption{\centering The correlation between the longitudinal charge density and the polar angle.}
		\label{fig:longitudinalChargeDensity}
	\end{minipage}
	\hfill
	\begin{minipage}{0.49\linewidth}
		\includegraphics[width=\linewidth]{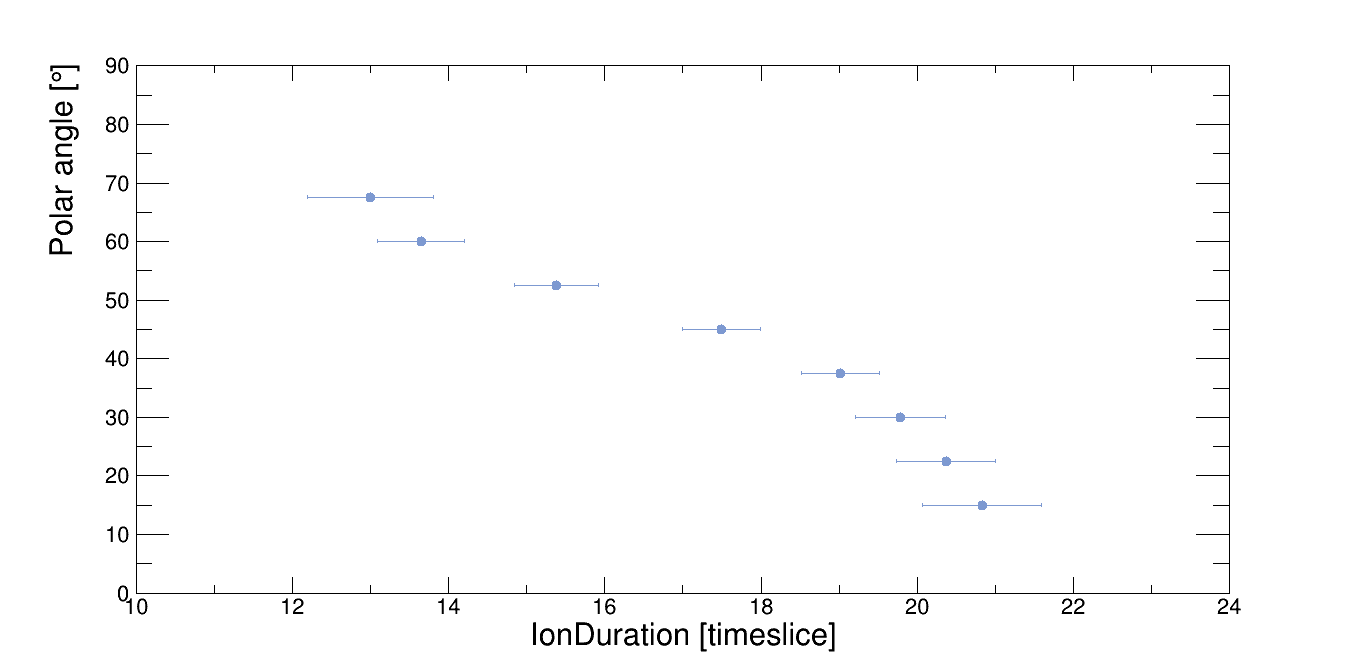}
		\subcaption{\centering The correlation between the polar angle and \texttt{IonDuration}.}
		\label{fig:thetaVsDuration}
	\end{minipage}
	\caption{Monte-Carlo simulations showing the influence of the polar angle on the longitudinal charge density and on \texttt{IonDuration} after $12.5~\rm{cm}$ of drift. The simulations model $10^4$ proton recoils induced by $27~\rm{keV}$ neutrons and the error bars correspond to the statistical uncertainties at one sigma.}
\end{figure*}
	
	We here describe a novel approach for accessing directionality (\textit{i.e.} measuring simultaneously $E_p$ and $\theta$) from the deconvolution of the ionic contribution on the Flash signal. The neutron field produced by the AMANDE facility covers a detection solid angle that can be determined for each event from the pixelated anode, with a maximal deviation of the neutron incident direction to the Z-axis that lies below $6^\circ$ (for a distance of $33~\mathrm{cm}$ between the target and the detector). In this situation, the scattering angle is roughly equivalent to the polar angle. Since we ignore the absolute Z-coordinate of the interaction, we place it (in our analysis) at the center of the detector, leading to a maximal error of $1.7^\circ$ in the worst-case scenario.
	
There is a direct correlation between the polar angle and the detector sensitivity to the ionic signal. To describe this correlation, we proceed in two steps. First, we remind that the larger the number of charges per timeslice at grid, the greater the ionic current, so the longer the ionic signal overpasses the detection threshold. We note \texttt{IonDuration} the duration between the time of arrival of the last primary electron and the time for which the Flash stops to record a signal. For instance in Figure~\ref{fig:sigFormation}, \texttt{IonDuration} is equal to $0$ for the brown curve, and $\sim 200\,\mathrm{ns}$ for the blue curve. The second step consists in determining the number of charges per timeslice at grid, a quantity that we hereafter call the \textit{longitudinal charge density}. Three main processes influence this quantity, all of them being related to the polar angle: (1) the energy transferred to the proton recoil by the neutron; (2) the track length projected along the Z-axis; (3) the diffusion of the charges in their drift towards the grid, whose deviation to the mean value is mainly observed, statistically, for a small number of charges, so for a large polar angle. One can show that the longitudinal charge density, and consequently \texttt{IonDuration}, decreases when the polar angle increases. In a first approximation, \texttt{IonDuration} and the polar angle are correlated by an inverse cosine function.

To confirm these tendencies, we have implemented a Monte-Carlo simulation that generates $n$ initial primary electrons clouds placed at the center of the detector ($z = 12.5\,\mathrm{cm}$) and that drifts them towards the grid. In this simulation, the scattering angle $\theta$ is fixed and it enables the determination of the proton kinetic energy, $E_p$, from Eq.~(\ref{eq:recoilE}) with $E_n=27\,\mathrm{keV}$; $E_p$ is then converted into ionization energy using our measured IQF; the primary electrons clouds are obtained from \texttt{SRIM}; and the diffusion coefficients and drift velocity are retrieved from \texttt{Magboltz}. In other words, we model $n$ proton recoils induced by elastic collisions with $27\,\mathrm{keV}$ neutrons in the detector. The correlation between the longitudinal charge density and the polar angle is presented in Figure~\ref{fig:longitudinalChargeDensity}. The next step uses the Dris and Alexopoulos model to determine the Flash signals induced by the simulated clouds. We finally apply our deconvolution and we consequently determine \texttt{IonDuration}. The results are presented in Figure~\ref{fig:thetaVsDuration}. As expected, the simulations show a correlation between \texttt{IonDuration} and the polar angle that can be described, in a first approximation, by an inverse cosine function. The simplicity of this function allows to easily adapt it to each experimental working condition.

\begin{figure*}
	\centering
	\begin{minipage}{0.49\linewidth}
		\includegraphics[width=1.1\linewidth]{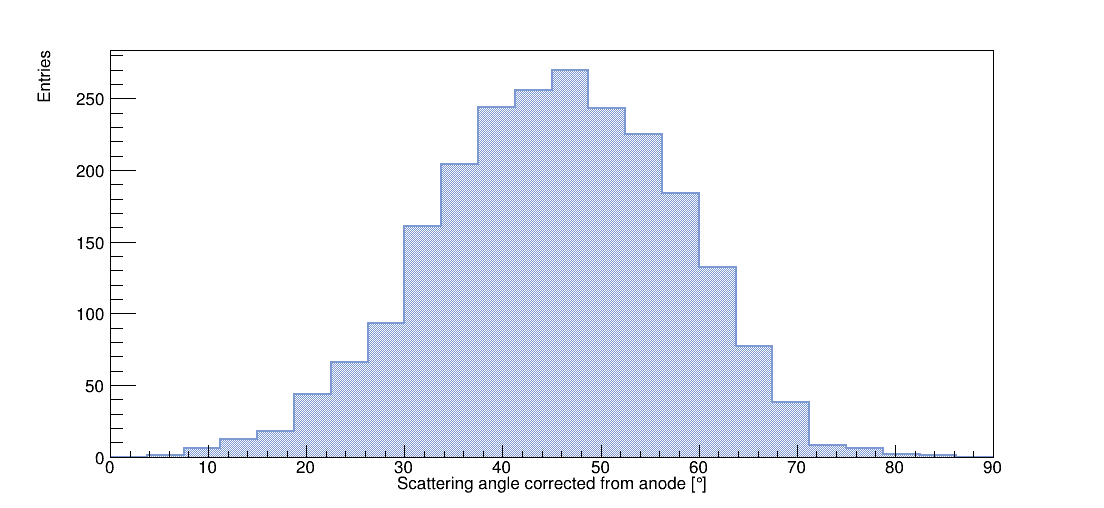}
		\subcaption{Angular distribution}
		\label{fig:angularDistrib27keV}
	\end{minipage}
	\hfill
	\begin{minipage}{0.49\linewidth}
		\includegraphics[width=1.1\linewidth]{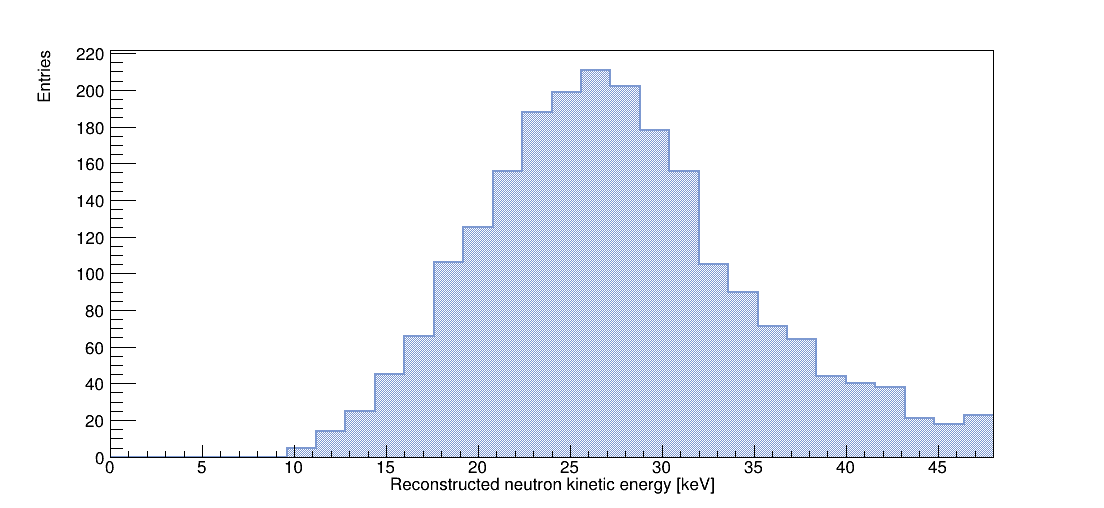}
		\subcaption{Energy spectrum}
		\label{fig:spectrum27keV}
	\end{minipage}
	\caption{Reconstruction of the neutron scattering angle distribution and the energy spectrum of a mono-energetic neutron field at $27\,\mathrm{keV}$.}
	\label{fig:27keV}
	
	\vspace*{0.2cm}
	\includegraphics[width=0.75\linewidth]{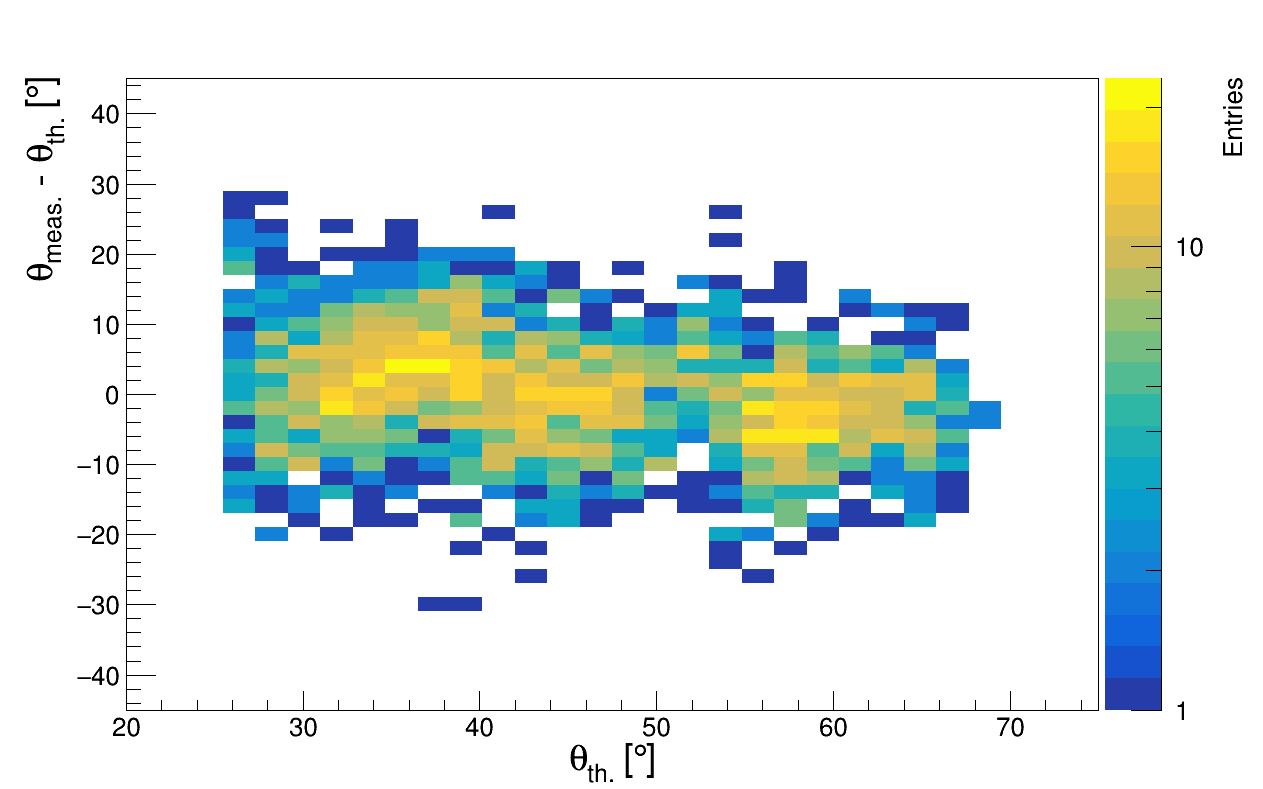}
	\caption{Angular resolution of the reconstruction of the neutron spectrum at $27\,\mathrm{keV}$.}
	\label{fig:angularResolution27keV}
\end{figure*}

The measurement of \texttt{IonDuration} (obtained from the deconvolution of the Flash signal) provides an indirect determination of the scattering angle for a neutron field aligned with the Z-axis of the detector. This procedure only relies on the Flash signal that has a low detection threshold (lower than the anode's one) and is consequently sensitive to low-energy recoils. We also emphasize that this method gets more precise when the gain increases with direct advantages for searches dedicated to the low-energy region. Finally, this approach based on the Flash signal can be coupled with the standard method of 3D track reconstruction from the anode. It then offers a redundancy on the measurement of the polar angle.

	\subsection{Reconstruction of mono-energetic neutron spectra at $27\,\mathrm{keV}$ and $8\,\mathrm{keV}$}

We apply the procedure described above to determine the scattering angle from the Flash signal in order to reconstruct the kinetic energy of the incident neutron with Eq.~(\ref{eq:recoilE}). The reconstructed spectrum at $27\,\mathrm{keV}$, presented in Figure~\ref{fig:27keV}, shows a Gaussian-like tendency with a resolution of $56\%$ (FWHM over mean value) and it peaks at $26.3\,\mathrm{keV}$, so we under-estimate the energy by $4\%$. The energy spectrum reconstruction from the measurement of the scattering angle embeds all experimental uncertainties introduced by the directional method (by deconvolution of the ionic signal), but also by the IQF, the energy calibration, and the electron-recoil discrimination. Note also that we only consider proton recoils, for simplicity, neglecting consequently the $13\%$ of carbon and fluorine recoils.

The performances of the angle reconstruction are presented in Figure~\ref{fig:angularResolution27keV} where $\theta_{meas.}$ is the reconstructed angle and $\theta_{th.}$ is the theoretical angle determined from Eq.~(\ref{eq:recoilE}) by setting $E_n=27.24\,\mathrm{keV}$. No significant bias is observed on the mean value of $\theta_{meas.} - \theta_{th.}$ and the standard deviation remains below $12^\circ$ in the entire range. This standard deviation corresponds to the angular resolution. Two competing phenomena explain the uniform behaviour of the angular resolution: on the one hand, a low scattering angle results in a large number of charges (according to Eq.~(\ref{eq:recoilE})) which improves the track resolution. On the other hand, the number of charges per timeslice increases for large scattering angles, leading to a better sensitivity to the ionic signal (\textit{c.f.} Figure~\ref{fig:thetaVsDuration}). We mention that the detector was saturating for kinetic energies above $22~\mathrm{keV}$, explaining why no event is detected for theoretical angles below $26^\circ$ in Figure~\ref{fig:angularResolution27keV}. Such saturation does not limit our analysis since proton recoils above $22~\mathrm{keV}$ are statistically rare.

\begin{figure*}
	\centering
	\begin{minipage}{0.49\linewidth}
		\includegraphics[width=1.1\linewidth]{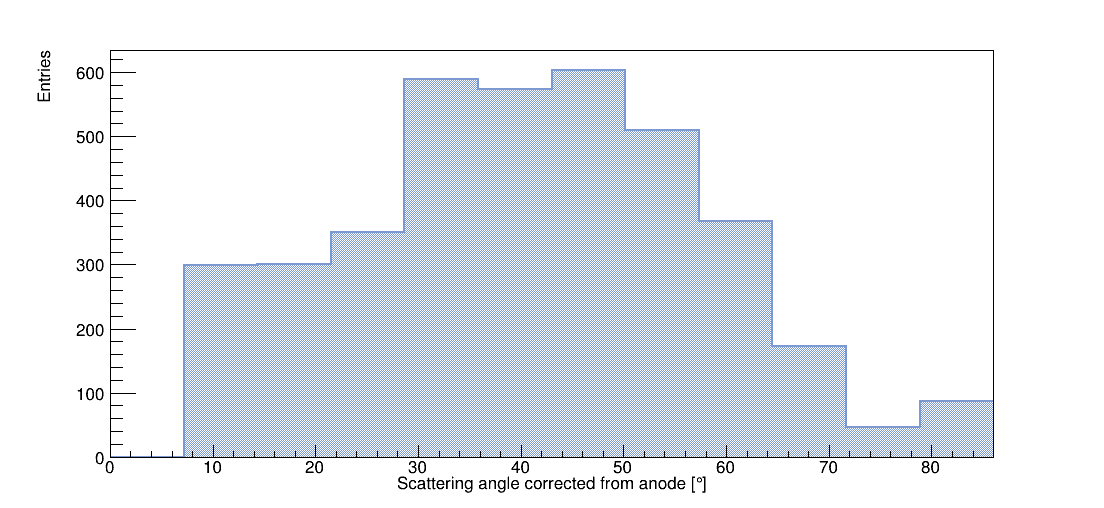}
		\subcaption{Angular distribution}
		\label{fig:angularDistrib8keV}
	\end{minipage}
	\hfill
	\begin{minipage}{0.49\linewidth}
		\includegraphics[width=1.1\linewidth]{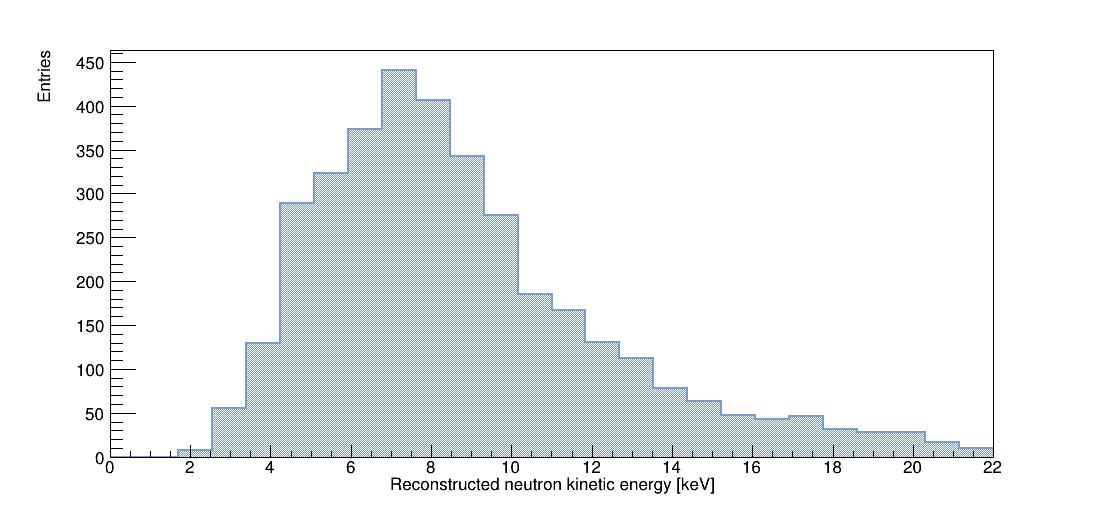}
		\subcaption{Energy spectrum}
		\label{fig:spectrum8keV}
	\end{minipage}
	\caption{Reconstruction of the neutron scattering angle and the energy spectrum at $8\,\mathrm{keV}$.}
	\label{fig:8keV}
	
	\vspace*{0.2cm}
	\includegraphics[width=0.75\linewidth]{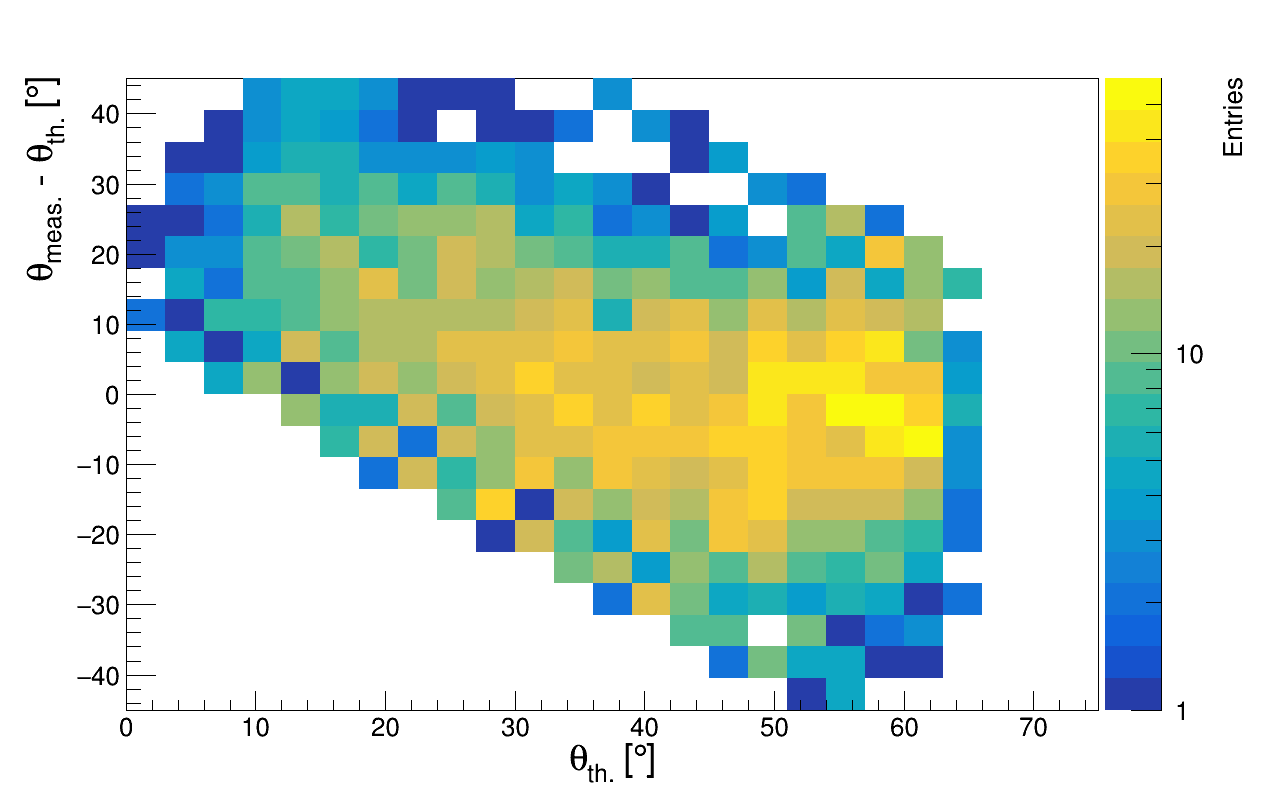}
	\caption{Reconstruction of the neutron scattering angle distribution and the energy spectrum of a mono-energetic neutron field at $8\,\mathrm{keV}$.}
	\label{fig:angularResolution8keV}
\end{figure*}

The analysis at $8\,\mathrm{keV}$ is presented in Figures \ref{fig:8keV} and \ref{fig:angularResolution8keV}. We reconstruct a neutron spectrum peaking at $7.4\,\mathrm{keV}$, \textit{i.e.} $9\%$ less than expected, the difference being explained by the presence of background events at low energy where the electron-recoil discrimination efficiency decreases. This phenomenon is also observed in Figure~\ref{fig:angularResolution8keV} where the low scattering angle region (below $15^\circ$) is almost exclusively populated by background events. We nevertheless measure an angular resolution better than $15^\circ$ in the entire angular range. 

Several comments are required at this stage. First, the reconstructions of the neutron energy spectra suffer from experimental uncertainties but they nevertheless demonstrate the directional sensitivity of the MIMAC detector to proton recoils with ionization energy in the keV-range. To the best of our knowledge, this work presents the first reconstruction of a neutron spectrum below $27\,\mathrm{keV}$ in a TPC.

Second, the measured directional performances for proton recoils in the keV-range (the range required for WIMP searches) are good enough to fulfil the requirements for a directional detector. An angular resolution of $30^\circ$ should be sufficient to discriminate light WIMPs from solar neutrinos \cite{OHare2015}. An angular resolution of $20^\circ$ with sense recognition and no background contamination would lead to a $3\sigma$ sensitivity at $90\%$ C.L. down to $10^{-5}\,\mathrm{pb}$ for spin-dependent cross-sections in a $30~\rm{kg\cdot year}$ CF$_4$ experiment \cite{Billard2011}. 

Finally, the main comment concerns the strategy of directional detection. The standard MIMAC method relies on the pixelated anode to reconstruct the direction of the nuclear recoil. This strategy has enabled for instance to reconstruct a neutron spectrum at $27\,\mathrm{keV}$ when operating at low gain with a gap of $256\,\mathrm{\mu m}$ \cite{Maire2016}. In this paper, we highlight how the situation changes at high gain. We have proposed an approach for directionality from the deconvolution of the ionic contribution on the Flash signal. We focused on the Flash signal for its simplicity to demonstrate the importance of the ionic contribution and to develop deconvolution tools. The next step consists in applying a similar approach on the pixelated anode measurements. The deconvolution of the signal on the anode requires additional work since its weighting field is non-linear and the anode's strips retrieve binary information (fired or not) instead of a direct measurement of the induced charge on the strip. The results presented in this work are definitely promising: the deconvolution of the ionic signal on the pixelated anode should improve the MIMAC directional performances in the near future. The approach from the Flash deconvolution described in this work will offer complementarity and redundancy to the angle measurement from 3D track reconstruction from the pixelated anode. 

\section{Conclusion}

In this paper, we have studied the influence of the ions on the signal formation in a Micromegas through three complementary and nested approaches: simulations, analytical model, and experimental measurements. The frequent back and forth between these approaches iteratively led to a consistent description of the phenomenon. When operating at high gain, the detector gets more sensitive to the slow kinematics of the numerous ions produced in the avalanche while at lower gain it mainly detects the intense and abrupt electronic signal. Most of the measured observables, the energy aside, are affected by the charge density in the amplification region and depend on the detector gain. 

The implementation of a simulation tool based on this scenario, \texttt{SimuMimac}, yielded to a new interpretation of previously published results. \texttt{SimuMimac} has validated that the measured elongation of fluorine tracks down to $6.3\,\mathrm{keV}$, as well as the empirical correction initially proposed, could be explained by the contribution of the ionic signal. It also brought us to the important conclusion that the MIMAC measurements are not altered by an event-based space-charge effect in the presented operating experimental conditions. From this point, and relying on an analytical model for the signal formation in a Micromegas, we have proposed a deconvolution of the Flash signal (\textit{i.e.} the induced charge integrated over time) to extract the electronic current from the measurements. This method can be applied to any measurements with no prior nor \textit{ad hoc} parameter and has been validated on several experimental data. It gives access to the time distribution of the primary electrons cloud at the Micromegas grid and its integral is proportional to the energy. Once the slow ionic signal is separated from the electronic one, the fine structure of the primary electrons cloud appears and it reveals an asymmetry in the time distribution of the primary charges. This asymmetry, being related to the Bragg peak on nuclear recoil tracks, can be used for an event-by-event head-tail recognition.

Directional detection is the only known strategy, so far, able to overpass the neutrino floor and establish a WIMP detection without ambiguity. From this perspective, detectors must access directionality in the low-energy region (nuclear recoils in the keV-range) which requires operating at high gain to be sensitive to any individual primary charge. We have highlighted in this work the complexity of the signal formation in the high gain region and we have developed tools to face this issue. Besides opening the window for low-energy searches, the deconvolution of the Flash also offers a new degree of freedom for data analysis by comparing the detector sensitivity to the electronic and the ionic signals. We have pointed out how this interplay can be used to indirectly measure the scattering angle of an elastic neutron-proton collision in a MIMAC chamber when the position of the target producing the neutron field is known. Thanks to this approach, we reconstructed a neutron energy spectrum at $27\,\mathrm{keV}$ and at $8\,\mathrm{keV}$. While several experimental uncertainties propagate to the final results, we measure a better than $15^\circ$ angular resolution for all angles considered. 

At high gain, the influence of the ions blurs and distorts the track measurements that constitute the backbone of the directional detection strategy based on 3D track reconstruction. In this paper, we focused on the deconvolution of the ionic signal for demonstrating its importance at high gain in a Micromegas. This work, developed for the MIMAC detector, could be adapted to other directional detectors. Now that we have established and validated tools for describing the influence of the ions on the Flash signal at high gain, we aim to push them one step further in deconvolving the ionic signal on the pixelated anode. Such a deconvolution would release the full directional performances of the MIMAC detector. We expect then to have in the near future two complementary and independent approaches, from the Flash signal and the pixelated anode, to measure the direction of a DM-induced nuclear recoil in the keV-range. 
 
\acknowledgments{We are very grateful to Charling Tao and Yi Tao for the fruitful discussions about high gain measurements, the cornerstone of this work. We thank Marine Hervé for her help during the AMANDE irradiations. We want to thank Michaël Petit and Thibaut Vinchon from the IRSN Micro-irradiation, Neutron Metrology and Dosimetry laboratory (LMDN-IRSN) hosting the AMANDE facility for their help during the experiments.}

\newpage
\appendix
\section{Working principles of \texttt{SimuMimac}}\label{app:SimuMimac}

The code of \texttt{SimuMimac} is available on request to the authors. It is a 2D implementation that follows the characteristics of the MIMAC detector as detailed in Section \ref{sec:sigForm}. The code makes use of \texttt{Garfield++} in two ways: (1) to determine the gas properties; (2) to compute the electric fields as well as the anode strip weighting fields. 

The primary electrons cloud can be generated manually by creating a uniform or a Gaussian cloud at any position in the detector. This requires knowing the W-value of the gas, \textit{i.e.} the mean energy needed to form an electron-ion pair. Another approach consists in calling \texttt{TRIM} \cite{SRIM} inside the code in order to simulate the properties of the primary electrons cloud generated by a nuclear recoil. This situation implies relying on the Ionization Quenching Factor of \texttt{SRIM} which is underestimated in low pressures gas mixtures \cite{Guillaudin2012, Santos2008}.

The transport of the electrons (drift and avalanche) uses the same implementation than in \texttt{Garfield++}. We have however encountered some issues with \texttt{Garfield++} for the motion of ions below $100\,\mathrm{mbar}$, so we have implemented our own solver to transport the ions. The equations of motion for the ions are given by the Langevin equation:
\begin{equation}
	m\frac{d\vec{v}}{dt} = q\big(\vec{E} + \vec{v}\times\vec{B}\big) - \frac{q}{\mu}\,\vec{v}
	\label{eq:langevin}
\end{equation}
where $\mu$ is the ion mobility that depends on the ion, the gas, the electric field, and the pressure. The Langevin equation is solved numerically by a Runge-Kutta-Fehlberg (4,5) method. Here we encounter two difficulties. First, several ion species are produced during the avalanche \cite{Torres2002}. Second, the mobilities of the ions are mainly unknown at the reduced electric fields considered (between $800\,\mathrm{Td}$ and $1300\,\mathrm{Td}$ depending on the gas mixtures used). We make the approximation of a mean mass with a mean mobility. In other words, we assume that a single ion type is produced during the avalanche and that it drifts with a constant mobility that does not depend on the electric field. We choose the mass of the ion with the largest ionization cross-section (for instance CF$_3^+$ in CF$_4$ \cite{Torres2002}) and we estimate a mobility value from an extrapolation of available data (for example using the \texttt{LXCat} database \cite{LXCAT}) in the closest experimental conditions. The mobility value is finally adjusted manually to match track depth measurements as explained in Section \ref{sec:SimuMimac}. This approximation limits the accuracy of \texttt{SimuMimac} and reminds us that such simulations must be used as a tool for investigating physical cases but that they always must be accompanied by measurements for experimental validation. 

We compute the signal induced on the grid and on the anode strips thanks to the Ramo-Shockley theorem, Eq.~(\ref{eq:Ramo}). We follow the approach of \cite{Dris2014} that considers a uniform weighting field for the grid. The weighting field of each strip is more complex and we use \texttt{Garfield++} to determine them. Part of the secondary ions can escape the amplification area and enter the drift region \cite{Colas2004}. This phenomenon is called \textit{ion backflow} and our simulations estimate it to lie between 2\% and 3\% of the total ions produced in the avalanche, depending on the applied voltages. Such ions induce a current that lies 3 orders of magnitude below the one from the ions in the amplification area. For this reason, we track the backflow ions only up to a small distance ($500\,\mathrm{\mu m}$) away from the grid. The simulations stop when there is no longer charge to transport and it outputs the current and the charge induced on the grid and on the strips of the anode.

\subsection*{Local distortions of the electromagnetic field}

In the first implementation of \texttt{SimuMimac} we were computing the electromagnetic (EM) field induced by the motion of the ions the gap (we assume an instantaneous drift of the electrons so we did not include them in the calculation). It represented the mainspring of writing our own simulation code. The local distortions of the EM field are then taken into account to transport the particles. As explained in Section \ref{sec:SimuMimac}, taking into account this possible space-charge effect did not influence the simulations results so we have eventually removed the implementation in the code. Anyway, we here give a short description of how it worked.

The root of the simulation of the space-charge effect is to loop over time. Before transporting a particle, we compute the local distortions of the EM fields by considering the properties of the closest ions. The usual approach for computing the induced EM field is to make use of Finite Element Method \cite{Bohmer2012}. However, this method would be too time-consuming in our situation since we would need a fine meshing to take the grid wire geometry into account and since we have to compute the field at any timestep. We instead decided to compute analytically the EM field. Due to the superposition principle, the EM field is the sum of the uniform field of the Micromegas ($\vec{F}^u$) plus the field induced by the ions ($\vec{F}^i$):
\begin{equation}
	\vec{F}(\vec{x}, t) = \vec{F}^u(\vec{x}, t) + \sum_{k=ions} \vec{F}_k^i(\vec{x}, t)
	\label{eq:superpos}
\end{equation}
where $\vec{F}$ can either be the electric field or the magnetic field. The EM field induced by a slowly moving charge can be computed in a covariant formalism. Under the approximation of a charge moving with constant velocity (which we assume in between two timesteps), the expression simplifies into \cite{Jackson1999}:

\begin{align}
	\vec{E}(\vec{x}, t) &= \frac{q}{4\pi\epsilon_0}\,\bigg( \frac{\vec{n} - \bm{\beta}}{\gamma^2\,\big(1 - \bm{\beta}\cdot\vec{n}\big)^3\,R^2}\bigg)_{ret} \nonumber\\
	\vec{B}(\vec{x}, t) &= \frac{1}{c}\,\bigg(\vec{n}\times\vec{E}\bigg)_{ret}
	\label{eq:EM}
\end{align}
where $\vec{x}$ is the position where we compute the electromagnetic field, $\bm{\beta} = \vec{v}(\tau)/c$ is the reduced velocity of the particle, $\tau$  is its proper time, $\gamma$ is the Lorentz factor, $\vec{n}$ is a unit vector in the direction $\vec{x} - \vec{r}(\tau)$, and $\vec{r}$ is the position of the particle. Finally, the index $_{ret}$ means that the expression must be evaluated at the retarded time $\tau_0$ for which the light-cone condition is fulfilled : $x_0 - r_0(\tau_0) = |\vec{x} - \vec{r}(\tau_0)| \equiv R$. The retarded time $\tau_0$ embeds the notion of the trajectories in the computation of the EM field. In the code, we decide that this time corresponds to the previous timestep. In other words, if we compute the EM field at the time $t_i$, then $\tau_0 = t_{i-1}$.

It would be too demanding for the code to take into account all charges in the chamber to compute the fields. Since the fields evolve as $1/r^2$ we only need to consider the closest neighbor charges. We do so by selecting a small area of the chamber in which we make the calculations. This area is cut into small pieces of about 1 $\mu$m width, each of them having a distinct number as an identifier (a \textit{key}). When computing the EM field at a given position, we only consider the charges located in the closest keys. Numerically, this approach is implemented by placing the ions in a C++ \texttt{unordered\_multimap}.


\section{Derivation of the deconvolution of the ionic signal}\label{app:deconvolution}

We use the analytical model of Dris and Alexopoulos \cite{Dris2014} to determine the electronic current from the Flash signal. The measured Flash corresponds to the integral of the charge induced on the grid over a timeslice of duration $\Delta t$: 
\begin{equation}
	C(t) ~=~ C(t-\Delta t) + \int_{t-\Delta t}^{t}d\tau \, \rho(\tau)\,\big(f \circledast g\big)(\tau)
\end{equation}
where $\rho(\tau)$ describes the charge density of the primary electrons cloud at grid, and where $f(\tau)$ and $g(\tau)$ are respectively the electronic and the ionic current given in Eq.~(\ref{eq:Dris}). We note $D(t)$ the charge difference in between two timeslices. We will here proceed by making an ansatz on the distribution $\rho(\tau)$: we assume a constant charge distribution, \textit{i.e.} $\rho(\tau) \simeq \rho = \frac{1}{N}$ where $N$ is a normalisation factor. This ansatz is a strong approximation, but it will enable us to perform analytically the deconvolution of the signal. The validity of the deconvolution is experimentally tested in Section \ref{sec:Prim}. With such ansatz, one has:
\begin{equation}
	D(t) ~\equiv~  C(t) - C(t - \Delta t) ~\simeq~ \frac{1}{N} \int_{t-\Delta t}^{t}d\tau \big(f \circledast g\big)(\tau)
\end{equation}
We will now make use of some properties of the Laplace transform. In the following, $s$ is the Laplace frequency associated with time in real domain. 
\begin{equation}
\mathcal{L}\big\lbrace D \big\rbrace ~=~ \frac{1}{Ns} \, \mathcal{L}\big\lbrace f\circledast g \big\rbrace ~=~ \frac{1}{Ns}  \mathcal{L}\big\lbrace f \big\rbrace \cdot \mathcal{L}\big\lbrace g \big\rbrace
\label{eq:Laplace}
\end{equation}
We can compute the Laplace transform of the ionic current:
\begin{align}
	\mathcal{L}\Big\lbrace g(t)\Big\rbrace(s) &= \frac{qu_p}{d}\int_0^\infty\,dt\bigg(e^{\alpha d} - e^{\alpha u_pt}\bigg)e^{-st}\nonumber\\
	&=\frac{qu_p}{d}\,\bigg(\frac{e^{\alpha d}}{s} - \frac{1}{s - \alpha u_p}\bigg)\nonumber\\
	&\simeq \frac{qu_p}{ds}\bigg(e^{\alpha d} -1\bigg)\,\bigg(1 - \frac{\alpha u_p}{s\big(e^{\alpha d} - 1\big)}\bigg)
\end{align}
where we have imposed the condition $s \gg \alpha u_p$ to Taylor expand the expression. This same condition brings us to the determination of the inverse:
\begin{equation}
	\frac{1}{\mathcal{L}\Big\lbrace g(t)\Big\rbrace(s)} \simeq  \frac{d}{qu_p\big(e^{\alpha d} -1\big)}\,\bigg(s + \frac{\alpha u_p}{e^{\alpha d} - 1}\bigg) \,=\,\frac{s+A}{B}
	\label{eq:1/L}
\end{equation}
We have introduced the parameters $A$ and $B$ to simplify the expressions:
\begin{equation}
	\begin{cases}
		A ~\equiv~ \frac{\alpha u_p}{e^{\alpha d} -1}\\
		B ~\equiv~ \frac{q u_p\big(e^{\alpha d} -1 \big)}{d}
	\end{cases}
	\label{eq:defAB}
\end{equation}
We now have all elements to inverse Eq.~(\ref{eq:Laplace}) and to return in the real time domain:
\begin{align}
	f(t) &~=~ \frac{N}{B}\,\mathcal{L}^{-1}\Big\lbrace s(s+A)\, \mathcal{L}\big\lbrace D(t) \big\rbrace \Big\rbrace\nonumber\\
	&~=~ \frac{N}{B}\,\Big( D''(t) + A\,D'(t)\Big)
\end{align}
where at the last line we used the fact that there is no charge at $t=0$. This is a differential equation of second order in $D(t)$, whose solution can be expressed as:
\begin{equation}
	D(t) ~=~ \int_0^tdx\bigg(  C_1e^{-Ax} + e^{-Ax}\int_0^x\,dy\,\frac{B}{N}e^{Ay}f(y) \bigg) + C_2
	\label{eq:integral}
\end{equation}
We can show that $C_1 = C_2 = 0$ since the Flash starts to record a signal before the arrival of the first primary electrons. Finally, we must apply a discretization since the Flash is a digital signal. Eq.~(\ref{eq:integral}) turns to:
\begin{align}
	D(t_i) &= \frac{B}{N}\,\sum_{j=0}^i\,\Delta t\,e^{-At_j}\,\sum_{k=0}^j\,\Delta t\,e^{At_k}\,f(t_k)\nonumber\\
	&= D(t_{i-1}) + \frac{B}{N}\,\Delta t\,e^{-At_i}\,\sum_{k=0}^i\,\Delta t\,e^{At_k}\,f(t_k)
\end{align}
We are now ready to express the electronic current as a function of the Flash signal:
\begin{equation}
	f(t_i) ~=~ \frac{N}{B\Delta t^2}\,\bigg( D(t_i) - \big(1 + e^{-A\Delta t}\big)D(t_{i-1}) + e^{-A\Delta t}\,D(t_{i-2})  \bigg)
	\label{eq:fCharge}
\end{equation}
This expression depends on physical quantities that cannot be easily determined experimentally. However, we are rather interested in the time distribution of the electronic current instead of its absolute value. For this reason, we can ignore the constant factor $\frac{N}{B\Delta t^2}$. The last step consists in determining $A$.

\subsection*{Experimental determination of $A$}

We would like to express $A$ from experimental Flash signals without computing it from the theoretical expression, Eq.~(\ref{eq:defAB}), since some physical quantities are unknown, and since it would require converting the charge into ADC-channel thanks to a calibration.

We evaluate $f(t)$ at two different timeslices, $t_\alpha$ and $t_\beta$. We can then extract the exponential term:

\begin{equation}
	e^{-A\Delta t} ~=~ \frac{\frac{f(t_\beta)}{f(t_\alpha)} \Big( D(t_\alpha) - D(t_{\alpha-1})\Big) - D(t_\beta) + D(t_{\beta-1})}{\frac{f(t_\beta)}{f(t_\alpha)} \Big( D(t_{\alpha-1}) - D(t_{\alpha-2})\Big) - D(t_{\beta-1}) + D(t_{\beta-2})}
\end{equation} 
The next step is to determine the ratio $f(t_\beta)/f(t_\alpha)$ from the Flash signal. There are two typical positions in the Flash that are always identifiable in the data: (1) the first detected charges; (2) the maximum of the Flash derivative. While the first case is mainly correlated with the electronic current, the second one depends both on the electronic and the ionic contributions. We will anyway proceed with the following approximation:
\begin{equation}
	\frac{f(t_\beta)}{f(t_\alpha)} \simeq \frac{D(t_\beta)}{D(t_\alpha)}
	\label{eq:Aexp}
\end{equation}
Under this approximation, we can retrieve the electronic current (in arbitrary units since we ignore the proportional factor in Eq.~(\ref{eq:fCharge})) from any Flash signal, without introducing any physical quantity nor \textit{ad hoc} parameters. In other words, the deconvolution can be applied to all measurements in any experimental condition. The evaluation of the performances of the deconvolution is detailed in Section \ref{sec:Prim}.

\bibliographystyle{JHEP}
\bibliography{biblio}

\end{document}